\documentclass[pdflatex,sn-nature]{sn-jnl}

\usepackage{graphicx}%
\usepackage{multirow}%
\usepackage{amsmath,amssymb,amsfonts}%
\usepackage{amsthm}%
\usepackage{mathrsfs}%
\usepackage[title]{appendix}%
\usepackage{xcolor}%
\usepackage{textcomp}%
\usepackage{manyfoot}%
\usepackage{booktabs}%
\usepackage{algorithm}%
\usepackage{algorithmicx}%
\usepackage{algpseudocode}%
\usepackage{listings}%
\usepackage{hyperref}

\usepackage{enumitem}
\usepackage{geometry}

\renewcommand{\Comment}[1]{// #1}

\theoremstyle{thmstyleone}%
\theoremstyle{thmstyletwo}%

\theoremstyle{thmstylethree}%

\begin{document}
\newgeometry{top=2.1cm, bottom=2.1cm, left=3.5cm, right=3.5cm}

\title[Article Title]{A general optimization solver based on OP-to-MaxSAT reduction}


\author[1]{\fnm{Yuxin} \sur{Zhao}}\email{yuxinzhaozyx@163.com}

\author*[2]{\fnm{Han} \sur{Huang}}\email{huangh985@mail.sysu.edu.cn}

\author[3]{\fnm{Zhifeng} \sur{Hao}}\email{haozhifeng@stu.edu.cn}

\affil*[1]{\orgdiv{School of Software Engineering}, \orgname{South China University of Technology}, \orgaddress{\city{Guangzhou}, \postcode{510006}, \country{China}}}

\affil[2]{\orgdiv{School of Software Engineering}, \orgname{Sun Yat-sen University}, \orgaddress{\city{Zhuhai}, \postcode{519082}, \country{China}}}

\affil[3]{\orgdiv{Department of Mathematics, College of Science}, \orgname{Shantou University}, \orgaddress{\city{Shantou}, \postcode{515063}, \country{China}}}


\abstract{Optimization problems are fundamental in diverse fields, such as engineering, economics, and scientific computing. However, current algorithms are mostly designed for specific problem types and exhibit limited generality in solving multiple types of optimization problems. To enhance generality, we propose an automated reduction method named OP-to-MaxSAT reduction and a general optimization solver based on OP-to-MaxSAT reduction (GORED). GORED unifies the solving of multiple types of optimization problems by reducing the problems from optimization problems to MaxSAT instances in polynomial time and solving them using the state-of-the-art MaxSAT solver. The generality and solution quality of GORED are validated through experiments on 136 instances across 11 types of optimization problems. Experimental results demonstrate that GORED not only successfully solves a wide range of optimization problems but also yields solutions comparable in quality to those from existing methods, with no statistically significant differences observed. By introducing automated reduction, this work shifts the paradigm of optimization solvers from designing specialized algorithms for each problem type to employing a single algorithm for diverse problems. As a result, advances in this single algorithm can now drive progress in a wide range of optimization problems across various domains.}

\keywords{Optimization Problem, MaxSAT, Solver, Reduction, Generality}



\maketitle

\restoregeometry
\newgeometry{top=2.1cm, bottom=2.1cm, left=3cm, right=3cm}

\section{Introduction}\label{sec1}

Optimization problems (OP) are prevalent in various fields, such as engineering design, resource allocation, machine learning, and transportation. The primary goal of optimization problems is to maximize or minimize a specific objective function under given constraints. As real-world problems become more and more diverse across various fields, the optimization problem needs to introduce various complex constraints for specific application scenarios.
For example, in transportation, to meet the requirements of multi-level distribution, crowdshipping, and electric vehicle charging, vehicle routing problems (VRP) must consider not only traditional path optimization but also factors like multi-level vehicles\cite{huangEmbeddedHamiltonianGraphguided2022, wangHybridFuzzyCmeans2025}, multiple depots\cite{zouTwoindividualEvolutionaryAlgorithm2025, mehlawatHybridIntelligentApproach2020}, heterogeneous capacities\cite{fengSolvingGeneralizedVehicle2021, bertsimasOptimizingSchoolsStart2019} and charging station locations\cite{zouTwoindividualEvolutionaryAlgorithm2025, xingBilevelGraphReinforcement2023, yinWassersteinDistributionallyRobust2025}. In industrial production, job-shop scheduling problem (JSP) has evolved into multiple variants to handle requirements such as maintenance\cite{anMultiobjectiveFlexibleJobshop2023, suSelforganizingNeuralScheduler2023}, outsourcing\cite{suSelforganizingNeuralScheduler2023}, energy savings\cite{liCoevolutionDeepReinforcement2024, panBipopulationEvolutionaryAlgorithm2022}, and dynamic jobs\cite{anMultiobjectiveFlexibleJobshop2023, liuDynamicJobshopScheduling2024, leiLargeScaleDynamicScheduling2024}. The increasing diversity of requirements significantly raises the complexity of optimization problems.
If all types of optimization problems could be solved using a general method, it could not only reduce manual intervention on algorithm development but also promote knowledge integration across disciplines, accelerating the development and application of optimization technologies in various fields.

Despite significant progress in solving optimization problems, existing research still shows clear limitations in terms of generality. The mainstream methods can be broadly categorized into two categories: mathematical programming methods and heuristic methods.
Mathematical programming methods, such as linear programming\cite{kuangAcceleratePresolveLargescale2025, moEnergyefficientTrainScheduling2020}, integer programming\cite{wangCustomizedAugmentedLagrangian2024, dashApplianceLoadDisaggregation2021, ShenFreeEnergyMachine2025}, and quadratic programming\cite{lamFastBinaryQuadratic2022, shirMultiobjectiveMixedintegerQuadratic2025}, offer notable advantages in theoretical rigor and solution quality. However, they rely heavily on strict mathematical structures, such as convex optimization or linear constraints. These methods struggle with non-standard constraints like nonlinear constraints. Non-standard optimization problems require manual transformation into equivalent standard forms before mathematical programming methods can solve them.
Heuristic methods, such as genetic algorithms\cite{duanRobustMultiobjectiveOptimization2022, awadGeneticAlgorithmGA2023, liElasticStrategybasedAdaptive2023, krolApplicationGeneticAlgorithm2022, liuManyobjectiveJobshopScheduling2023, heMultiobjectiveOptimizationEnergyefficient2022}, evolutionary algorithms\cite{liTwopopulationAlgorithmLargescale2025, huangCorrelationbasedDynamicAllocation2024, yangLocaldiversityEvaluationAssignment2023, hanExploringRepresentationsOptimizing2024, gaoSolvingFuzzyJobshop2020}, and particle swarm optimization\cite{dziwinskiNewHybridParticle2020, daiMultistageParticleSwarm2025, hanNovelSetbasedDiscrete2024, zouariPSObasedAdaptiveHierarchical2022, zhangPromotiveParticleSwarm2022}, provide some flexibility. They allow designing operators for specific problems to adapt to the problem characteristics. However, this customization leads to high costs for algorithm transplantation and makes it challenging to create a general solver. Moreover, heuristic methods lack theoretical guarantees of optimality and are prone to getting stuck in local optima. Designing local search strategies for avoiding local optima further reduces their general applicability.
Both methods depend heavily on manual intervention, such as problem transformation and algorithm design. This reliance makes it hard to achieve the universal solving of optimization problems and handle increasingly complex optimization problems. Therefore, developing a general optimization method that can adapt to various types of optimization problems while reducing manual intervention is an important research direction.

To address the limitations of existing optimization methods in terms of generality, in this paper, we propose a general optimization solver based on OP-to-MaxSAT reduction (GORED). The core idea is to achieve universal optimization by automatically reducing any type of optimization problem to a MaxSAT instance and solving it using a MaxSAT solver. This optimization solver analyzes the constraint structure of problems and applies the proposed reduction rules to automate the reduction process, eliminating the need for manual intervention. We also demonstrate that the time complexity of this reduction process is polynomial, ensuring both efficiency and generality of the reduction. By introducing automated reduction, this work marks a shift from developing problem-specific algorithms to using a single algorithm for diverse optimization problems, enabling progress in one algorithm to generalize across many problem types.

In summary, this paper provides the following contributions: 
\begin{enumerate}[label=\arabic*)]
    \item We propose an algorithm called OP-to-MaxSAT reduction, which automatically reduces multiple types of optimization problems to MaxSAT instances in polynomial time. Through rigorous theoretical analysis, we prove its time complexity. This reduction ensures the reliability and efficiency of the general optimization solver when handling various optimization problems.
    \item We develop a general optimization solver based on the proposed OP-to-MaxSAT reduction. This solver achieves generality by automating the reduction and solving processes, enabling a single algorithm to solve multiple types of optimization problems, such as integer/mixed-integer programming, linear/nonlinear programming, combinatorial/numerical optimization, etc. This solver serves as a general tool for solving optimization problems across various fields.
    \item To validate the cross-field applicability and practical effectiveness of the proposed general optimization solver, we conducted experiments on 11 types of optimization problems from various fields. These experiments cover several real-world applications, such as transportation and industrial production. The results show that the solver can effectively address optimization challenges across fields, highlighting its potential and value in practical applications.
\end{enumerate}

\section{OP-to-MaxSAT Reduction}

The Maximum Satisfiability problem (MaxSAT) is an extension of the Boolean Satisfiability problem (SAT). A MaxSAT instance is defined as $maxcnf = (S, H)$ over a set of Boolean variables $X=\{x_1, x_2,\cdots,x_n\}$, where $S = \{ (C_1, w_1), (C_2,w_2),\cdots,(C_p,w_p) \}$ is a set of weighted soft clauses with $w_i \ge 0$, and $H = \{ C_{p+1}, C_{p+2}, \cdots,C_{p+q}\}$ is a set of hard clauses. Each clause $C_i$ is a disjunction of literals, where a literal is a Boolean variable $x_i$ or its negative $\neg x_i$. For example, $C_1=x_1 \lor \neg x_2 \lor x_5$. A clause is satisfied if at least one of its literals is true under a given variable assignment $V: X \to \{0, 1\}$. A hard clause is a clause that must be satisfied. MaxSAT aims to find a solution (variable assignment) $V$ that maximizes the total weight of satisfied soft clauses in $S$, while satisfying all hard clauses in $H$. This structure is highly similar to optimization problems, making MaxSAT an ideal target problem for reduction from optimization problems.

Although many studies have attempted to manually reduce specific types of optimization problems to MaxSAT instances, these methods are limited to specific types of problems and lack generality\cite{leiSolvingSetCover2020, ciampiconiMaxSATbasedFrameworkGroup2020, molaviQubitMappingRouting2022}. Such manual reductions are not only time-consuming and labor-intensive, but also difficult to adapt to new types of optimization problems. To perform a manual reduction, one must first understand the characteristics of the specific problem before designing the corresponding reduction rules, which greatly limits the applicability and practicality of these methods. Therefore, developing an automated and general reduction method is the key to overcoming these limitations.

To overcome the above limitations, we propose an automated reduction method called OP-to-MaxSAT reduction. This method consists of four steps: (1) unified modeling of optimization problems; (2) reducing variables in optimization problems to Boolean variables in MaxSAT; (3) reducing constraints to hard clauses; and (4) reducing the objective to soft clauses. This method not only improves the automation of reduction but also enhances its applicability across various optimization problems. In the following section, we will discuss the details of these steps.

\subsection{Unified Modeling of Optimization Problems}

To reduce various types of optimization problems to MaxSAT problems, a unified modeling language is needed to describe these optimization problems. Since LaTeX is widely used in academic communication, we have designed the modeling language for optimization problems in LaTeX's mathematical format. This format makes it easier to represent and visualize the mathematical models of optimization problems. Detailed grammar of this modeling language can be found in Appendix~\ref{appendix:grammar}.
The proposed modeling language formulates the optimization problem in the form of Formulas~\ref{formula:general-objective}-\ref{formula:general-constraint-n}.
\begin{align}
\min / \max \quad & O \label{formula:general-objective} \\
s.t. \quad & C_1 \quad B_1 \\
& C_2 \quad B_2 \\
& \space \vdots \quad \quad \vdots \\
& C_n \quad B_n \label{formula:general-constraint-n} 
\end{align}
where the objective of the problem consists of the optimization direction $\min / \max$ and the objective function $O$. The objective function $O$ is represented as a numerical expression. The $i$-th constraint of the problem consists of a constraint formula $C_i$ and a universal quantifier $B_i$. The constraint formula $C_i$ is a relational expression. The universal quantifier $B_i$ is optional, which defines the scope of bound variables in the constraint formula $C_i$.

In the following, we will use the Capacitated Vehicle Routing Problem (CVRP) as an example to illustrate the structure of optimization problems that the proposed modeling language can express in detail, as shown in Formulas~\ref{formula:model1}-\ref{formula:model12}.

\begin{align}
\min && & \sum_{k=1}^m \sum_{i=1}^n \sum_{j=1}^n d_{i,j} x_{i,j,k} \label{formula:model1} \\
s.t. && & x_{i,j,k} \in \{0,1\} && \forall k \in V, i,j \in P \label{formula:model2} \\
&& & x_{i,i,k} = 0 && \forall k \in V, i\in P \label{formula:model3} \\
&& & \sum_{i=1}^n x_{i,j,k} = \sum_{i=1}^n x_{j,i,k} && \forall j \in P, k\in V \label{formula:model4} \\
&& & \sum_{k=1}^m \sum_{i=1}^n x_{i,j,k} = 1 && \forall j \in C \label{formula:model5} \\
&& & \sum_{j=2}^n x_{1,j,k} = 1 && \forall k \in V \label{formula:model6} \\
&& & \sum_{i=1}^n \sum_{j=2}^n q_j x_{i,j,k} \le Q && \forall k \in V \label{formula:model7} \\
&& & u_j - u_i - q_j + Q \ge Q \max_{k=1}^m \{ x_{i,j,k} \} && \forall i,j \in C, \space i\neq j \label{formula:model8} \\
&& & u_i \in \{q_i, \dots, Q\} && \forall i \in C \label{formula:model9} \\
&& & V = \{1, \dots, m\} \label{formula:model10} \\
&& & C = \{2, \dots, n\} \label{formula:model11} \\
&& & P = \{1\} \cup C \label{formula:model12}
\end{align}

Formulas~\ref{formula:model1}-\ref{formula:model12} present a mathematical model of CVRP using the proposed modeling language. $m$, $n$, $d_{i,j}$, $q_j$ and $Q$ are the parameters of this problem. $m$ is the number of vehicles, $n$ is the number of positions, $d_{i,j}$ is the distance between position $i$ and position $j$, $q_j$ is the demand of customer $j$, and $Q$ is the capacity of each vehicle. In a problem instance, these parameters are defined after Formulas~\ref{formula:model1}-\ref{formula:model12} in the form of an assignment expression, such as $m=5$, $n=20$, $d_{1,2} = 3$, etc. Formula~\ref{formula:model1} defines the objective of the problem, including the optimization direction (max or min) and the expression of the objective function. Formulas~\ref{formula:model2}-\ref{formula:model12} are constraints of the problem. Each constraint consists of a constraint formula and a universal quantifier. A universal quantifier, such as $i,j\in C, i \neq j$ in Formula~\ref{formula:model8}, defines the scope of bound variables (such as $i,j$) in the constraint formula. A constraint formula is a relational expression. It can be classified into equality constraints (e.g., Formulas~\ref{formula:model3}-\ref{formula:model6}), inequality constraints (e.g., Formulas~\ref{formula:model2}, \ref{formula:model7}, \ref{formula:model8}), and domain constraints (e.g., Formulas~\ref{formula:model2}, \ref{formula:model9}-\ref{formula:model12}). A domain constraint specifies the type and domain of a variable. For example, $u_i \in \{q_i,\dots,Q\}$ indicates that $u_i$ is an integer variable ranging from $q_i$ to $Q$, and $x \in \mathbb{R}$ indicates that $x$ is a real variable. The ability to specify integer and real variables means that the modeling language is able to express mixed integer programming problems. In this modeling language, each expression supports not only basic arithmetic operations but also nonlinear operations such as power, absolute values, minimum, and maximum (e.g., Formula~\ref{formula:model8}). Therefore, this modeling language allows nonlinear constraints and objective, making it suitable for nonlinear optimization problems. The modeling language also supports set variables and set operations, as shown in Formulas~\ref{formula:model10}-\ref{formula:model12}. However, the purpose of supporting set variables is to simplify the problem description. Therefore, set variables cannot serve as decision variables in optimization problems. Instead, set variables can only be used as constant sets to represent finite integer or enumerated domains.

OP-to-MaxSAT reduction includes an iterative simplification process for this modeling language, shown in Algorithm~\ref{alg:op-to-maxsat-reduction}. This process aims to gradually eliminate all universal quantifiers and set variables, in order to simplify constraints and identify the set of decision variables. For example, from the constraint $m=5$, we can derive $V=\{1,\dots,5\}$, which allows Formula~\ref{formula:model6} to be split into five constraints without universal quantifiers. At the same time, abstract variables like $x_{1,j,k}$  can be instantiated into specific variables with concrete indices, such as $x_{1,2,1}$. Constraints such as $m=5$ and $V=\{1,\dots,5\}$ are treated as assignment expressions. All the assignment expressions are removed from the mathematical model during simplification. When no more constraints can be simplified, the variables remaining in the mathematical model form the set of decision variables of the optimization problem. This simplification process not only checks the validity of the input mathematical model, but also provides a simpler representation of the problem for reduction to a MaxSAT instance.

\subsection{Reduction of Variables}

To reduce optimization problems to MaxSAT instances, the numerical variables of optimization problems must be encoded into Boolean variables used in MaxSAT. In optimization problems, variables can be classified into integer and real variables, corresponding to integer programming and mixed-integer programming problems, respectively. To ensure the general applicability of variable encoding, we use the signed binary fixed-point representation for encoding variables.
\begin{align}
\boldsymbol a=\underbrace{a_{m+n}}_{\text{sign bit}} \underbrace{a_{m+n-1} \cdots a_{m}}_{\text{integer bits}} \underbrace{a_{m-1} \cdots a_{0}}_{\text{fractional bits}}
\label{formula:encoding-variable}
\end{align}
As shown in Formula~\ref{formula:encoding-variable}, the variable $\boldsymbol{a}$ can be encoded to $m+n+1$ Boolean variables, including 1 sign bit, $n$ integer bits, and $m$ fractional bits. The value it represents is $v(\boldsymbol a)= (-1)^{a_{m+n}} \sum_{-m}^{n-1} 2^i a_{i+m}$. For integer variables, all fractional bits can be set to 0 to exactly represent integer values without loss of precision. For real number variables, although this encoding can only approximate real values, the precision can be flexibly controlled by adjusting the number of fractional bits $m$, to meet the requirements of specific problems. This encoding method not only accurately represents numerical variables but also facilitates the reduction of constraints and objective in later steps.

\subsection{Reduction of Constraints}

Given a solution to an optimization problem, checking whether it satisfies all constraints is an NP problem. According to the theory of NP-completeness, any NP problem can be reduced to a SAT problem in polynomial time. Therefore, the constraints of an optimization problem can be reduced to a SAT instance, i.e., a conjunctive normal form (CNF). This CNF can be treated as hard clauses in the MaxSAT problem, ensuring that any feasible solution to the MaxSAT problem strictly satisfies all constraints of the original problem. If no MaxSAT solution satisfies these hard clauses, then the original optimization problem has no feasible solution.
\begin{figure}[h]
\centering
\includegraphics[height=2.2cm]{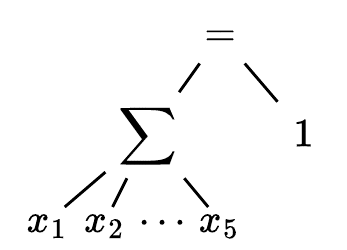}
\caption{Operation tree of constraint $\sum_{i=1}^5 x_i = 1$}\label{fig:operation-tree}
\end{figure}

{
\setlength{\tabcolsep}{3pt}
\begin{table}[b]
\centering
\caption{Reduction Rules of Operations}
\label{tab:reduction-rules-of-operations}
\begin{tabular}{ccl}
\toprule
Category & Operator & Reduction Rule \\
\midrule
Relational & $=$ & $Equal(\boldsymbol a, \boldsymbol b)$ \\
Operations & $\neq$ & $NotEqual(\boldsymbol a, \boldsymbol b)$ \\
       & $\le, \ge$ & $LessEqual(\boldsymbol a, \boldsymbol b)$ \\
       & $<, >$ & $LessThan(\boldsymbol a, \boldsymbol b)$ \\
\midrule
Arithmetic & $+, \Sigma$ & $Sum(\{\boldsymbol a_1, \boldsymbol a_2, \cdots, \boldsymbol a_k\}, \boldsymbol c)$ \\
Operations & $\times,\cdot,\Pi$ & $Product(\{\boldsymbol a_1, \boldsymbol a_2, \cdots, \boldsymbol a_k\}, \boldsymbol c)$ \\
       & $k \boldsymbol a$ & $Scale(\boldsymbol a, k, \boldsymbol c)$ \\
       & $\boldsymbol a^k, /$ & $Power(\boldsymbol a, k, \boldsymbol c)$ \\
       & $\|\boldsymbol a\|$ & $Absolute(\boldsymbol a, \boldsymbol c)$ \\
       & $\lfloor \boldsymbol a \rfloor$ & $Floor(\boldsymbol a, \boldsymbol c)$ \\
       & $\lceil \boldsymbol a \rceil$ & $Ceil(\boldsymbol a, \boldsymbol c)$ \\
       & $\max$ & $Max(\{\boldsymbol a_1, \boldsymbol a_2, \cdots, \boldsymbol a_k\}, \boldsymbol c)$ \\
       & $\min$ & $Min(\{\boldsymbol a_1, \boldsymbol a_2, \cdots, \boldsymbol a_k\}, \boldsymbol c)$ \\
\midrule
Domain & $\in$ & $IntegerDomain(\boldsymbol a, L, R)$ \\
Definitions & $\in$ & $EnumerationDomain(\boldsymbol a, \{\boldsymbol b_1, \boldsymbol b_2, \cdots, \boldsymbol b_k\})$ \\
       & $L < \boldsymbol a \le R$ & $RealDomain(\boldsymbol a, L_1, R_1, L_2, R_2)$ \\
\bottomrule
\end{tabular}
\vspace{-1cm}
\end{table}
}

After the simplification process of the modeling language, each constraint in the mathematical model of the optimization problem is a relational expression. Each constraint can be represented as an operation tree, as shown in Figure~\ref{fig:operation-tree}. In the operation tree, each intermediate node represents an operation, and each leaf node represents a variable or a constant. Based on the properties of operations, these operations can be divided into three categories: relational operations, arithmetic operations, and domain definitions. For each category of operation, we have designed a corresponding reduction rule to reduce it to CNF. Table~\ref{tab:reduction-rules-of-operations} summarizes the correspondence between operations and reduction rules. Detailed reduction rules are provided in Appendix~\ref{appendix:reduction-rules}. By reducing all nodes in the operation tree to CNF according to their corresponding reduction rules, all constraints can be converted to CNF.

The reduction rule for the relational operation $\boldsymbol a \operatorname{op_r} \boldsymbol b$ can be uniformly expressed as $OP_r(\boldsymbol a, \boldsymbol b)$, where $\operatorname{op_r}$ is a relational operator, $\boldsymbol a$ and $\boldsymbol b$ are two operands of this operation. In the proposed modeling language, each relational operation corresponds to the relational expression in a constraint. Since all such expressions must be true, these reduction rules do not need to encode the result of relational operations. Instead, they only need to describe how these two operands make the relational operation true. For example, the CNF generated by $Equal(\boldsymbol a, \boldsymbol b)$ is as Formulas~\ref{equal-rule-start}-\ref{equal-rule-end}.

\begin{align}
&& & \neg a_{m+n} \lor \neg b_{m+n} \lor e \label{equal-rule-start} \\
&& & a_{m+n} \lor b_{m+n} \lor e \\
&& & a_{m+n} \lor \neg b_{m+n} \lor \neg e \\
&& & \neg a_{m+n} \lor b_{m+n} \lor \neg e \\
&& & \neg e \lor \neg a_i \lor b_i && i=0,\dots,m+n-1 \\
&& & \neg e \lor a_i \lor \neg b_i && i=0,\dots,m+n-1 \\
&& & e \lor \neg a_i && i=0,\dots,m+n-1 \\
&& & e \lor \neg b_i && i=0,\dots,m+n-1 \label{equal-rule-end}
\end{align}
where $a_i$ and $b_i$ are the binary bits of the operands $\boldsymbol{a}$ and $\boldsymbol{b}$ under signed binary fixed-point encoding; these bits serve as the propositional variables of SAT problems. $e$ is an additional intermediate variable introduced for representing this operation.

The reduction rule for the arithmetic operation $\boldsymbol c = \operatorname{op_a} (\boldsymbol a_1, \dots, \boldsymbol a_k)$ can be uniformly represented as $OP_a(\{\boldsymbol a_1, \cdots, \boldsymbol a_k\}, \boldsymbol c)$. Here, $\operatorname{op_a}$ denotes the function performing the arithmetic operation, $\boldsymbol a_1$ to $\boldsymbol a_k$ are its $k$ operands, and $\boldsymbol c$ is the output of the operation. The operands $\boldsymbol a_1$ to $\boldsymbol a_k$ can be obtained from the children of the operation node, while the output $\boldsymbol c$ is an intermediate variable created specifically for this operation node. The purpose of such reduction rules is to describe the relationship between operands and output.

The reduction rule for the domain definition $\boldsymbol a \in D$ can be uniformly represented as $OP_d(\boldsymbol a, D)$, where $D$ is the domain of the operand $\boldsymbol a$. We define $IntegerDomain$ and $RealDomain$ to describe continuous integer and real number domains, respectively. In addition, we define $EnumerationDomain$ to describe sparse and non-continuous domains.

\subsection{Reduction of the Objective}

In the proposed modeling language, the objective function is an arithmetic expression composed of arithmetic operations. Therefore, we first construct the operation tree of the objective function. Then, we encode all arithmetic operations in the objective function into hard clauses using the constraint reduction method described earlier. The output of the root node is taken as the variable representing the objective value, denoted by $\boldsymbol u$. After these steps, the only remaining task is to encode the objective value $\boldsymbol u$ into soft clauses of MaxSAT.

The goal of the optimization problem is to maximize or minimize the objective value, while the MaxSAT problem only supports maximization. To handle optimization problems that aim to minimize the objective value, we convert the minimization of the objective value into the maximization of its negation. For a signed binary fixed-point encoded objective value $\boldsymbol u$, we only need to flip the sign bit to obtain the unified maximization objective value $\boldsymbol \mu$, as shown in Formula~\ref{formula:objective-flip}.
\begin{align} \label{formula:objective-flip}
\boldsymbol \mu = \begin{cases}
u_{m+n}u_{m+n-1}\cdots u_{0}, & \text{maximize } \boldsymbol u \\
(\neg u_{m+n})u_{m+n-1}\cdots u_{0}, & \text{minimize } \boldsymbol u \\
\end{cases}
\end{align}

For a signed objective value $\boldsymbol \mu$, if treated as a signed integer, its range is $[-2^{m+n}+1, 2^{m+n}-1]$. If we can shift it to the range $[1, 2^{m+n}-1]$, denoted by an unsigned integer $\boldsymbol {\hat \mu}$, $\hat \mu_0$ to $\hat \mu_{m+n}$ can be used as soft clauses in the MaxSAT problem. The weight of each soft clause $\hat \mu_i$ can be set to $2^i$, so that the objective function of the MaxSAT problem becomes $\sum_{i=0}^{m+n} 2^i \hat \mu_i$. This transformation realizes the reduction of the objective from the optimization problem to the MaxSAT problem. To convert the objective value $\boldsymbol \mu$ into $\boldsymbol {\hat \mu}$, a simple way is $\boldsymbol {\hat \mu} = \boldsymbol \mu + 2^{m+n}$. However, the value $2^{m+n}$ exceeds the range of an $(m+n+1)$-bit signed number. To avoid overflow, we define the following reduction rule $Normalization(\boldsymbol \mu, \boldsymbol {\hat \mu})$:
\begin{align}
    Complement(\boldsymbol \mu, (\neg \hat \mu_{m+n}) \hat \mu_{m+n-1}\cdots \hat \mu_0)
\end{align}
where $Complement(\boldsymbol a, \boldsymbol c)$ indicates that $\boldsymbol c$ is the two's complement of $\boldsymbol a$. See Appendix~\ref{appendix:reduction-rules} for details. This reduction rule converts $\boldsymbol \mu$ into $\boldsymbol {\hat \mu}$ by taking the two's complement of $\boldsymbol \mu$ and then flipping its sign bit.

\subsection{Reduction Algorithm}

In summary, the reduction algorithm proposed in this paper includes the simplification process of the modeling language and the reduction processes of variables, constraints, and the objective, shown in Algorithms~\ref{alg:op-to-maxsat-reduction}-\ref{alg:encoding-of-expression}.

\begin{algorithm}[H]
\caption{OP-to-MaxSAT Reduction}
\label{alg:op-to-maxsat-reduction}
\begin{algorithmic}[1]
\Require $M$: optimization problem instance,
\Statex $n$: number of integer bits,
\Statex $m$: number of fractional bits
\Ensure $maxcnf$: generated MaxSAT instance,
\Statex $P$: set of known quantities and their values,
\Statex $Q$: set of unknown quantities and their encodings

\State $P = \emptyset, Q = \emptyset$
\State // simplify the mathematical model
\While{$M$ exists assignment expression} \Comment {equations such as $c = 1$}
    \State extract an assignment expression ``$variable = value$'' from $M$
    \State replace all ``$variable$'' of $M$ with ``$value$''
    \State perform algebraic simplification on $M$ 
    \State add $(variable, value)$ into $P$
\EndWhile
\State // reduction of constraints
\For{each constraint $C$ in $M$}
    \State represent the expression of $C$ as an operation tree $T$
    \State $H =$ \Call{EncodeExpression}{$T$, $Q$}
    \State add $H$ into $maxcnf$ as hard clauses
\EndFor
\State // reduction of the objective function
\State represent the expression of objective function in $M$ as an operation tree $T$
\State $H =$ \Call{EncodeExpression}{$T$, $Q$}
\State add $H$ into $maxcnf$ as hard clauses
\State $\boldsymbol u = $ the encoding of $T$'s root
\If{the optimization direction of $M$ is $\operatorname{minimize}$}
    \State $\boldsymbol \mu = (\neg u_{m+n}) u_{m+n-1} \cdots u_{0}$
\Else
    \State $\boldsymbol \mu = \boldsymbol u$
\EndIf
\State create a new intermediate variable $\boldsymbol {\hat \mu}$
\State add $Normalization(\boldsymbol \mu, \boldsymbol {\hat \mu})$ into $maxcnf$ as hard clauses
\For{$i=0$ to $m+n$}
    \State add $\hat \mu_{i}$ as soft clause and $2^i$ as its weight into $maxcnf$
\EndFor
\State \Return $maxcnf$, $P$, $Q$
\end{algorithmic}
\end{algorithm}

\begin{algorithm}[H]
\caption{Encoding of Expression}
\label{alg:encoding-of-expression}
\begin{algorithmic}[1]
\Require
\Statex $T$: the operation tree of expression,
\Statex $Q$: set of variables and their encodings
\Ensure $H$: generated hard clauses
\Function{EncodeExpression}{$T$, $Q$}
    \State $H=\emptyset$
    \For{each node $e$ of $T$ in post order}
        \If{$e$ is a variable}
            \If{$e$ is in $Q$}
                \State the encoding of $e$ = the encoding of $e$ in $Q$
            \Else
                \State // reduction of variable
                \State encode variable $e$ as a signed binary fixed-point number $\boldsymbol c$
                \State the encoding of $e$ = $\boldsymbol c$
                \State add ($e$, $\boldsymbol c$) into $Q$
            \EndIf
        \ElsIf{$e$ is a numeric constraint}
            \State encode the output of $e$ as a signed binary fixed-point number $\boldsymbol c$
            \State the encoding of $e$ = $\boldsymbol c$
        \ElsIf{$e$ is a relational operation}
            \State $\boldsymbol a,\boldsymbol b =$ the encodings of $e$'s two children
            \State add $OP_r(\boldsymbol a,\boldsymbol b)$ into $H$
        \ElsIf{$e$ is an arithmetic operation}
            \State encode the output of $e$ as a signed binary fixed-point number $\boldsymbol c$
            \State the encoding of $e$ = $\boldsymbol c$
            \State $\boldsymbol a_1, \cdots, \boldsymbol a_k$ = the encodings of $e$'s children
            \State add $OP_a(\{\boldsymbol a_1, \cdots, \boldsymbol a_k\}, \boldsymbol c)$ into $H$
        \ElsIf{$e$ is a domain definition}
            \State $D = $ the domain specified in $e$
            \State $\boldsymbol a = $ the variable defined in $e$
            \State add $OP_d(\boldsymbol a, D)$ into $H$ 
        \EndIf
    \EndFor
    \State \Return $H$
\EndFunction
\end{algorithmic}
\end{algorithm}

\newpage

\section{Results}

\begin{figure}[h]
\centering
\includegraphics[width=15cm]{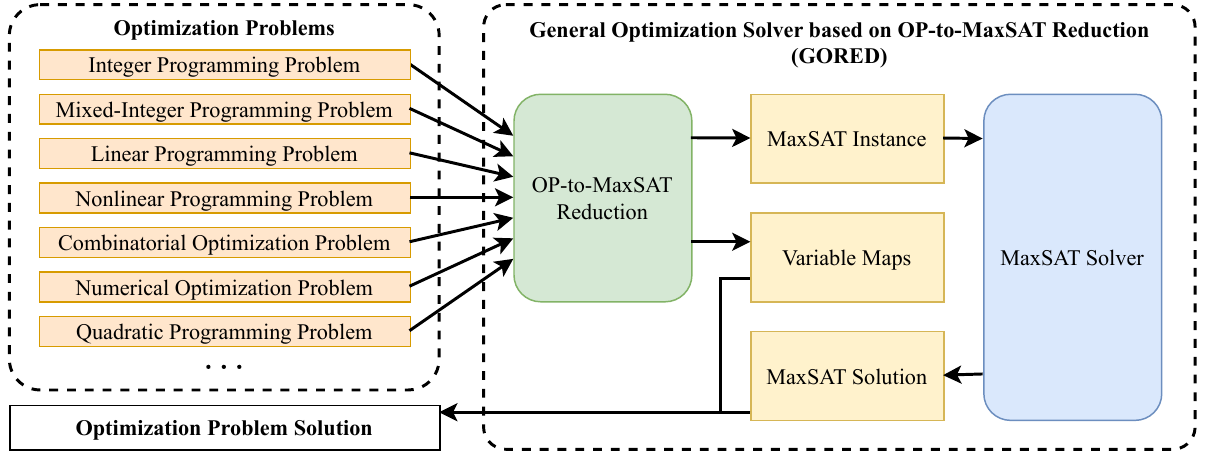}
\caption{The Architecture of General Optimization Solver based on OP-to-MaxSAT Reduction}\label{fig:GORED-arch}
\end{figure}

The scientific viewpoint of this paper is that automated reduction is the key to enabling multiple types of optimization problems to be solved by a single algorithm. To validate this viewpoint, this paper proposes GORED, a general optimization solver enabled by the OP-to-MaxSAT reduction. Its architecture is shown in Figure~\ref{fig:GORED-arch}. GORED reduces input optimization problems into MaxSAT instances, employs standard MaxSAT solvers to solve these MaxSAT instances, and finally maps the MaxSAT solutions back to the original problem space to obtain high-quality solutions for the original optimization problems. Without incorporating problem-specific solving logic, GORED enables unified modeling and solving of multiple types of optimization problems, shifting the paradigm of optimization solvers from designing specialized algorithms for each problem type to employing a single algorithm for a wide range of optimization problems.

\subsection{Generality}

To verify the generality of GORED, we conducted experiments on 11 optimization problems. The mathematical models of these problems, as represented by the proposed modeling language, are shown in Appendix~\ref{appendix:models}. These problems cover typical types of optimization problems, including integer programming, mixed-integer programming, linear programming, nonlinear programming, numerical optimization, combinatorial optimization, quadratic programming, convex optimization, and non-convex optimization. The application scenarios span multiple fields, such as transportation and industrial production. In Table~\ref{tab:comparsion-in-generality}, we compare the manual intervention required by GORED and mainstream optimization methods to adapt to these problems. The compared mainstream optimization methods include mathematical programming solvers like CPLEX, Gurobi, and SCIP, as well as heuristic methods like genetic algorithm(GA), evolutionary algorithm(EA), and particle swarm optimization(PSO). The experimental results show that mainstream optimization methods require manual intervention to adapt to specific optimization problems, while GORED can solve various types of problems without manual intervention. In Table~\ref{tab:comparsion-in-generality-algorithm}, we compare the solving algorithms used by GORED and mainstream optimization methods for each problem. The table shows that mainstream methods employ different solving algorithms for different problems, whereas GORED can uniformly solve these problems using just a single algorithm. The results indicate that GORED has stronger generality in terms of manual intervention and solving algorithms.

{
\begin{sidewaystable}
    \centering
    \begin{minipage}{0.48\textwidth}

\centering
\caption{Comparison in Generality: Manual Intervention}
\label{tab:comparsion-in-generality}
\begin{tabular}{l c c c c c c c c c c c}
\toprule
Optimization Problem & \multicolumn{4}{c}{Problem Characteristics} & \multicolumn{7}{c}{Manual Intervention} \\
\cmidrule(lr){2-5} \cmidrule(lr){6-12}
& LP/NLP & IP/MIP & NOP/COP & others & GORED & CPLEX & Gurobi & SCIP & GA & EA & PSO \\
\midrule
Graph Coloring Problem\cite{BeasleyORLibrary1990}                 & NLP & IP  & COP  &   -         & O & O & R & R & R+S   & E+R & E+R  \\
Traveling Salesman Problem\cite{gerhardTSPLIB1991}                & LP  & IP  & COP  &   -         & O & O & O & O & E+R+S & E+R & E+R  \\
Capacitated Vehicle Routing Problem\cite{uchoaCVRPLIB2017}        & LP  & MIP & COP  &   -         & O & O & O & R & E+R+S & E+R & E+R  \\
Two-Echelon Vehicle Routing Problem\cite{BeasleyORLibrary1990}    & LP  & MIP & COP  &   -         & O & O & O & O & E+R+D & E+R & E+R  \\
Multidimensional Knapsack Problem\cite{BeasleyORLibrary1990}      & LP  & IP  & COP  &   -         & O & O & O & O & S     & E   & E    \\
Job-Shop Scheduling Problem\cite{TAILLARD1993}                    & LP  & MIP & COP  &   -         & O & O & O & O & E+R+D & E+R & E+R  \\
Open-Shop Scheduling Problem\cite{TAILLARD1993}                   & LP  & MIP & COP  &   -         & O & O & O & O & E+R+D & E+R & E+R  \\
Quadratic Assignment Problem\cite{QAPLIB1991}                     & NLP & IP  & COP  & QP          & O & O & O & O & E+S   & E   & E  \\
Shifted Sphere Function\cite{CEC2005}                             & NLP & MIP & NOP  & QP + convex & O & O & O & O & S     & O   & O    \\
Shifted Schwefel's Problem\cite{CEC2005}                          & NLP & MIP & NOP  & QP + convex & O & O & O & O & S     & O   & O    \\
Shifted Rosenbrock's Function\cite{CEC2005}                       & NLP & MIP & NOP  & non-convex  & O & X & R & R & S     & O   & O    \\
\bottomrule
\end{tabular}
\footnotetext{Legend: LP = linear programming, NLP = nonlinear programming, IP = integer programming, MIP = mixed-integer programming, NOP = numerical optimization problem, COP = combinatorial optimization problem, QP = quadratic programming, convex = convex optimization problem, non-convex = non-convex optimization problem, O = no manual intervention need, X = unable to adapt to this problem, R = reformulating problem, E = encoding decision variables, S = selecting proper operators, D = designing problem-specific operators.}
    
    \end{minipage}
    
    \hfill \vspace{1cm}
    
    \begin{minipage}{0.48\textwidth}

\centering
\caption{Comparison in Generality: Solving Algorithms}
\label{tab:comparsion-in-generality-algorithm}
\begin{tabular}{l c c c c c c c}
\toprule
Optimization Problem & \multicolumn{7}{c}{Solving Algorithms} \\
\cmidrule(lr){2-8}
& GORED & CPLEX & Gurobi & SCIP & GA & EA & PSO \\
\midrule
Graph Coloring Problem\cite{BeasleyORLibrary1990}                 & MaxSAT Solver & B\&C    & B\&C           & B\&C         & GA+TPX+PM   & DE+Discretization & PSO+Discretization  \\
Traveling Salesman Problem\cite{gerhardTSPLIB1991}                & MaxSAT Solver & B\&C    & B\&C           & B\&C         & GA+OX+IM & DE+Discretization & PSO+Discretization  \\
Capacitated Vehicle Routing Problem\cite{uchoaCVRPLIB2017}        & MaxSAT Solver & B\&C    & B\&C           & B\&C         & GA+OX+IM & DE+Discretization & PSO+Discretization  \\
Two-Echelon Vehicle Routing Problem\cite{BeasleyORLibrary1990}    & MaxSAT Solver & B\&C    & B\&C           & B\&C         & GA+CX+CM & DE+Discretization & PSO+Discretization  \\
Multidimensional Knapsack Problem\cite{BeasleyORLibrary1990}      & MaxSAT Solver & B\&C    & B\&C           & B\&C         & GA+TPX+BM     & DE+Discretization   & PSO+Discretization    \\
Job-Shop Scheduling Problem\cite{TAILLARD1993}                    & MaxSAT Solver & B\&C    & Simplex+B\&C & B\&C         & GA+CX+CM & DE+Discretization & PSO+Discretization  \\
Open-Shop Scheduling Problem\cite{TAILLARD1993}                   & MaxSAT Solver & B\&C    & B\&C           & B\&C         & GA+CX+CM & DE+Discretization & PSO+Discretization  \\
Quadratic Assignment Problem\cite{QAPLIB1991}                     & MaxSAT Solver & B\&C    & B\&C           & B\&C         & GA+OX+IM     & DE+Discretization & PSO+Discretization  \\
Shifted Sphere Function\cite{CEC2005}                             & MaxSAT Solver & Barrier & Barrier        & B\&C         & GA+SBX+PM     & DE   & PSO    \\
Shifted Schwefel's Problem\cite{CEC2005}                          & MaxSAT Solver & Barrier & Barrier        & B\&C         & GA+SBX+PM     & DE   & PSO    \\
Shifted Rosenbrock's Function\cite{CEC2005}                       & MaxSAT Solver & -       & B\&C           & B\&C + Ipopt & GA+SBX+PM     & DE   & PSO    \\
\bottomrule
\end{tabular}
\footnotetext{Legend: B\&C = branch-and-cut, Ipopt = interior point optimizer, Simplex = simplex method, Barrier = barrier method, TPX = two-point crossover, OX = order crossover, SBX = simulated binary crossover, BM = bitflip mutation, IM = inversion mutation, PM = polynomial mutation, CX = custom problem-specific crossover, CM = custom problem-specific mutation, DE = differential evolution algorithm, Discretization = decoding continuous decision variables to discrete variables.}
    
    \end{minipage}
\end{sidewaystable}
}

\newpage

\subsection{Solution Quality}

To evaluate the performance of GORED in terms of solution quality, we conducted experiments on 136 test instances across 11 types of optimization problems described above. The results are summarized in Table~\ref{tab:comparison-in-solution-quality}, see Appendix~\ref{appendix:detailed-results} for details. The results indicate that GORED satisfies all constraints and achieves the optimal objective value in all problem instances, achieving solution quality comparable to current optimization methods. Its superiority mainly comes from the automated OP-to-MaxSAT reduction, i.e., automatically converting the original optimization problem into a MaxSAT instance and then solving it using a MaxSAT solver. This reduction makes the completeness of GORED depend on the completeness of the MaxSAT solver. When a complete MaxSAT solver is used, GORED is also complete, and the solution found is guaranteed to be the global optimum of the original optimization problem under finite precision.

{
\setlength{\tabcolsep}{3pt}
\begin{table}[h]
\centering
\caption{Comparison in Solution Quality}
\label{tab:comparison-in-solution-quality}
\begin{tabular}{l c c c c c c c c}
\toprule
Optimization Problem & \# Instances & \multicolumn{6}{c}{Comparison of GORED with Existing Methods ($+/=/-$)} \\
\cmidrule(lr){3-8}
& & CPLEX & Gurobi & SCIP & GA & EA & PSO \\
\midrule
Graph Coloring Problem\cite{BeasleyORLibrary1990}              & 13  & 0/13/0 & 0/13/0 & 0/13/0 & 12/1/0 & 12/1/0 & 13/0/0   \\
Traveling Salesman Problem\cite{gerhardTSPLIB1991}             & 11  & 0/11/0 & 0/11/0 & 0/11/0 & 11/0/0 & 11/0/0 & 11/0/0    \\
Capacitated Vehicle Routing Problem\cite{uchoaCVRPLIB2017}     &  8  & 0/8/0  & 0/8/0  & 0/8/0  & 2/6/0  & 7/1/0  & 8/0/0     \\
Two-Echelon Vehicle Routing Problem\cite{BeasleyORLibrary1990} & 20  & 0/20/0 & 0/20/0 & 0/20/0 & 19/1/0 & 0/20/0 & 16/4/0      \\
Multidimensional Knapsack Problem\cite{BeasleyORLibrary1990}   & 20  & 0/20/0 & 0/20/0 & 0/20/0 & 2/18/0 & 1/19/0 & 18/2/0     \\
Job-Shop Scheduling Problem\cite{TAILLARD1993}                 & 10  & 0/10/0 & 0/10/0 & 0/10/0 & 10/0/0 & 10/0/0 & 10/0/0      \\
Open-Shop Scheduling Problem\cite{TAILLARD1993}                & 20  & 0/20/0 & 0/20/0 & 0/20/0 & 10/10/0 & 10/10/0 & 20/0/0     \\
Quadratic Assignment Problem\cite{QAPLIB1991}                  &  8  & 0/8/0  & 0/8/0  & 0/8/0  & 1/7/0  & 0/8/0  & 1/7/0      \\
Shifted Sphere Function\cite{CEC2005}                          & 10  & 0/10/0 & 0/10/0 & 0/10/0 & 0/10/0 & 0/10/0 & 1/9/0      \\
Shifted Schwefel's Problem\cite{CEC2005}                       &  6  & 0/6/0  & 0/6/0  & 0/6/0  & 5/1/0  & 4/2/0  & 2/4/0    \\
Shifted Rosenbrock's Function\cite{CEC2005}                    & 10  &   -    & 0/10/0 & 0/10/0 & 10/0/0 & 9/1/0  & 10/0/0     \\
\midrule
Total                               & 136 & 0/126/0 & 0/136/0 & 0/136/0 & 82/54/0 & 64/72/0 & 110/26/0 \\
\bottomrule
\end{tabular}
\footnotetext{Legend: The notation "$+/=/-$" indicates the number of instances in which GORED performs better than / equally to / worse than the existing methods in terms of the objective value of solution.}
\end{table}
}

\subsection{Polynomial Time Complexity of Reduction}

To demonstrate the computational efficiency of the OP-to-MaxSAT reduction proposed in this paper, we analyze the time complexity of the entire reduction process. The total reduction time $T(P)$ for an optimization problem $P$, with $n$ integer bits and $m$ fractional bits, can be expressed as:
\begin{align}
T(P)=T_m(P)+T_c(P)+T_t(P)
\end{align}
where $T_m(P)$ denotes the time spent on the mathematical model simplification, $T_c(P)$ denotes the time for reducing constraints in problem $P$, and $T_t(P)$ denotes the time for reducing the objective in problem $P$. To analyze how the size of problem $P$ affects the reduction time, we use $p$, the number of operations, and $q$, the maximum number of operands among all operations, to represent the size of problem $P$.

During the simplification of the mathematical model, the OP-to-MaxSAT reduction selects one assignment expression at a time for simplification, until no assignment expressions remain. To find an assignment expression, each constraint must be checked. Since the number of constraints is less than the number of operations, the time to find one assignment expression is $O(p)$. There are $O(p)$ assignment expressions in total. Each time an assignment expression is found, OP-to-MaxSAT reduction replaces variables and performs algebraic simplification of operations. These involve traversing $O(pq)$ operands and $O(p)$ operations. Therefore, the total time spent on mathematical model simplification $T_l(P)$ is in $O(p^3 q)$.

In the proposed OP-to-MaxSAT reduction, reducing constraints means converting each operation into a conjunctive normal form (CNF) according to the corresponding reduction rule. Therefore, the reduction time for constraints can be expressed as the sum of the time required by each operation's reduction rule. The time needed for each reduction rule depends on the size of the generated CNF, i.e., the number of literals and clauses, as shown in Table~\ref{tab:reduction-rules-complexity}. The table shows that the maximum time complexity of a single reduction rule is $O(qm^2+qn^2+qmn)$. Thus, the total reduction time for constraints $T_c(P)$ is in $O(pqm^2+pqn^2+pqmn)$.
{
\setlength{\tabcolsep}{3pt}
\begin{table}[t]
\centering
\caption{Generated CNF Size and Time Complexity of Reduction Rules}
\label{tab:reduction-rules-complexity}
\begin{tabular}{l @{\hspace{-0.7cm}} c c c}
\toprule
Reduction Rule & \multicolumn{2}{c}{Size of The Generated CNF} & Time Complexity \\
\cmidrule(lr){2-3}
& Upper Bound of \#Literals & Upper Bound of \#Clauses &  \\
\midrule
$Equal(\boldsymbol a, \boldsymbol b)$
& $O(1)$
& $O(m+n)$ 
& $O(m+n)$ \\

$NotEqual(\boldsymbol a, \boldsymbol b)$ 
& $O(m+n)$ 
& $O(m+n)$ 
& $O(m+n)$ \\

$LessEqual(\boldsymbol a, \boldsymbol b)$ 
& $O(m+n)$
& $O(m+n)$
& $O(m+n)$ \\

$LessThan(\boldsymbol a, \boldsymbol b)$ 
& $O(m+n)$
& $O(m+n)$
& $O(m+n)$ \\

$Sum(\{\boldsymbol a_1, \boldsymbol a_2, \cdots, \boldsymbol a_k\}, \boldsymbol c)$ 
& $O(qm+qn)$
& $O(qm+qn)$
& $O(qm+qn)$ \\

$Product(\{\boldsymbol a_1, \boldsymbol a_2, \cdots, \boldsymbol a_k\}, \boldsymbol c)$ 
& $O(qm^2+qn^2+qmn)$ 
& $O(qm^2+qn^2+qmn)$ 
& $O(qm^2+qn^2+qmn)$ \\

$Scale(\boldsymbol a, k, \boldsymbol c)$ 
& $O(m^2 + n^2 + mn)$
& $O(m^2 + n^2 + mn)$
& $O(m^2 + n^2 + mn)$ \\

$Power(\boldsymbol a, k, \boldsymbol c)$ 
& $O(m^2+n^2+mn)$
& $O(m^2+n^2+mn)$
& $O(m^2+n^2+mn)$ \\

$Absolute(\boldsymbol a, \boldsymbol c)$ 
& $O(1)$
& $O(m+n)$
& $O(m+n)$ \\

$Floor(\boldsymbol a, \boldsymbol c)$ 
& $O(n)$
& $O(m+n)$
& $O(m+n)$ \\

$Ceil(\boldsymbol a, \boldsymbol c)$ 
& $O(n)$
& $O(m+n)$
& $O(m+n)$ \\

$Max(\{\boldsymbol a_1, \boldsymbol a_2, \cdots, \boldsymbol a_k\}, \boldsymbol c)$ 
& $O(qm+qn)$
& $O(qm+qn)$
& $O(qm+qn)$ \\

$Min(\{\boldsymbol a_1, \boldsymbol a_2, \cdots, \boldsymbol a_k\}, \boldsymbol c)$ 
& $O(qm+qn)$
& $O(qm+qn)$
& $O(qm+qn)$ \\

$IntegerDomin(\boldsymbol a, L, R)$ 
& $O(m+n)$
& $O(m+n)$
& $O(m+n)$ \\

$EnumerationDomain(\boldsymbol a, \{\boldsymbol b_1, \boldsymbol b_2, \cdots, \boldsymbol b_k\})$ 
& $O(qm+qn)$
& $O(qm+qn)$
& $O(qm+qn)$ \\

$RealDomin(\boldsymbol a, L_1, R_1, L_2, R_2)$ 
& $O(m+n)$
& $O(m+n)$
& $O(m+n)$ \\

$Normalization(\boldsymbol \mu, \boldsymbol {\hat \mu})$ 
& $O(m+n)$
& $O(m+n)$
& $O(m+n)$ \\
\bottomrule
\end{tabular}
\end{table}
}

The reduction of the objective mainly consists of two parts: encoding the arithmetic operations in the objective function and encoding the objective value. The encoding of the arithmetic operation is done in the same way as in the reduction of constraints, so the time spent in operation encoding is $O(pqm^2+pqn^2+pqmn)$. For encoding the objective value, we first apply the normalization rule $Normalization(\boldsymbol \mu, \boldsymbol {\hat \mu})$,  and then reduce it to $m+n+1$ soft clauses. As shown in Table~\ref{tab:reduction-rules-complexity}, the time complexity of normalization is $O(m+n)$. Hence, the time complexity for encoding the objective value is also $O(m+n)$. The total reduction time for objective $T_t(P)$ is in $O(pqm^2+pqn^2+pqmn)$.

In summary, the total time of OP-to-MaxSAT reduction $T(P)$ is in $O(p^3q + pqm^2 + pqn^2 + pqmn)$. This indicates that OP-to-MaxSAT runs in polynomial time. Since most optimization problems are NP-hard, and no polynomial-time algorithms are currently known for NP-hard problems, the time cost of OP-to-MaxSAT reduction only accounts for a small portion of the overall solving time for the optimization problem.

\section{Discussion}

The proposed GORED successfully addresses the challenge faced by traditional optimization methods in handling diverse and complex optimization problems with a single algorithm. By introducing an automated reduction method, GORED enables universal solving across various types of optimization problems, significantly reducing manual intervention. This breakthrough opens up new possibilities for cross-domain integration and application of optimization techniques. OP-to-MaxSAT reduction and GORED establish a new paradigm that solves multiple types of problems with a single algorithm, rather than designing specialized algorithms for each problem type, enabling advances in one algorithm to drive progress across a wide range of problems. To better understand the value and limitations of GORED, we will thoroughly discuss the shortcomings of existing optimization methods, analyze the advantages of GORED over them, and further discuss the limitations of GORED and potential directions for future improvement.

\newpage

\subsection{Comparison with Related Methods}
In contrast to existing methods, GORED introduces an automated reduction mechanism, allowing a single algorithm to solve diverse optimization problems while guaranteeing completeness.

Mathematical programming methods, such as branch-and-cut\cite{kazachkovAbstractModelBranch2024, galiasDeterministicBranchCut2025, castellucciNewBranchandbenderscutAlgorithm2025}, interior-point method\cite{zhangWarmstartStrategyInterior2023, luFullyDecentralizedOptimal2018, huangProactiveMobilityLoad2025}, barrier method\cite{dvurechenskyHessianBarrierAlgorithms2025, dzahiniConstrainedStochasticBlackbox2023, zhaoAnalysisFrankWolfe2023}, and simplex method\cite{blankGeneratingWellspacedPoints2021, xuScaledSimplexRepresentation2021}, are classic techniques in the field of optimization. They are known for their theoretical rigor and high solution quality. These methods classify optimization problems into many types. For example, linear programming (LP) is widely used in planning and scheduling\cite{kuangAcceleratePresolveLargescale2025, moEnergyefficientTrainScheduling2020}. Integer programming (IP) is used for combinatorial optimization\cite{wangCustomizedAugmentedLagrangian2024, dashApplianceLoadDisaggregation2021,ShenFreeEnergyMachine2025}. Quadratic programming (QP) is suitable for problems with a quadratic objective function\cite{lamFastBinaryQuadratic2022, shirMultiobjectiveMixedintegerQuadratic2025}. Each mathematical programming method is designed for a specific type of problem, which makes these methods rely on certain mathematical structures like convexity or linearity. They exhibit limitations when dealing with complex nonlinear constraints or objectives. In such cases, manual intervention is needed to relax constraints, linearize, or reformulate models. The manual intervention limits their generality and automation. With the increasing demand for solving optimization problems in industry, many optimization solvers have been developed by integrating multiple mathematical programming methods. Examples include CPLEX\cite{IBMILOGCPLEXOptimizationStudio2024}, Gurobi\cite{GUROBI2025}, and SCIP\cite{sureshbolusaniSCIPOptimizationSuite2024}.  These solvers offer unified modeling languages and support modeling and solving various optimization problems. Compared to a single mathematical programming method, they greatly extend the range of applicable problems. However, their ability to handle different problem types mainly depends on automatically selecting suitable mathematical programming methods according to the problem characteristics before solving the problems. In other words, their generality comes from integrating and packaging multiple algorithms, not from using a single algorithm to automatically solve multiple types of problems. Compared with these solvers,  GORED does not rely on the mathematical structure of problems or the integration of mathematical programming methods. Instead, it solves multiple types of problems using a single algorithm through OP-to-MaxSAT reduction. Therefore, GORED offers higher generality.

Heuristic methods are widely used in complex optimization problems due to their flexibility and adaptability. Typical approaches such as Genetic Algorithms (GA)\cite{duanRobustMultiobjectiveOptimization2022, awadGeneticAlgorithmGA2023, liElasticStrategybasedAdaptive2023, krolApplicationGeneticAlgorithm2022, liuManyobjectiveJobshopScheduling2023, heMultiobjectiveOptimizationEnergyefficient2022}, Evolutionary Algorithms (EA)\cite{liTwopopulationAlgorithmLargescale2025, huangCorrelationbasedDynamicAllocation2024, yangLocaldiversityEvaluationAssignment2023, hanExploringRepresentationsOptimizing2024, gaoSolvingFuzzyJobshop2020}, Particle Swarm Optimization (PSO)\cite{dziwinskiNewHybridParticle2020, daiMultistageParticleSwarm2025, hanNovelSetbasedDiscrete2024, zouariPSObasedAdaptiveHierarchical2022, zhangPromotiveParticleSwarm2022}, and Simulated Annealing (SA)\cite{davariOnlineWeightingFactor2021, onizawaFastconvergingSimulatedAnnealing2023, hanSimulatedAnnealingbasedHeuristic2023} have demonstrated effectiveness in diverse applications, including job-shop scheduling and vehicle routing. Recently, reinforcement learning\cite{mingConstrainedMultiobjectiveOptimization2024, pengLearningbasedTemporalSequence2025, yuanRLCSLCombinatorialOptimization2023, zhangMetalearningbasedDeepReinforcement2023, ghafarollahiModelingProteinMotions2024, barretoFastReinforcementLearning2020}, graph neural network\cite{heydaribeniDistributedConstrainedCombinatorial2024,SchuetzCombinatorialOptimizationwithPhysic}, and large language model\cite{akibaEvolutionaryOptimizationModel2025} have also been introduced into optimization, showing potential for generalization across various problem types. However, these methods generally lack theoretical guarantees of optimality. They are prone to getting stuck in local optima. Moreover, the design of heuristic algorithms is highly dependent on the specific problem. This problem-specific algorithmic design makes it difficult to use the same algorithm for different problems without modification, limiting their generality. In addition, parameter tuning and operator design of heuristic methods require substantial expert knowledge, which further limits their generality. Compared with heuristic methods, GORED does not rely on problem-specific operators. It can be applied to various types of optimization problems without algorithmic adjustments, offering higher generality. Besides, the completeness of GORED comes from the MaxSAT solver used after reduction. When a complete MaxSAT solver is used to solve the reduced MaxSAT instance, GORED can guarantee the optimality of solutions.

Reduction is not a new technique. Some studies have already attempted to reduce optimization problems to MaxSAT instances. These problems include the set cover problem\cite{leiSolvingSetCover2020}, the dominating set problem\cite{leiSolvingSetCover2020}, the group testing problem\cite{ciampiconiMaxSATbasedFrameworkGroup2020}, and more recently, the quantum bit mapping and routing problems\cite{molaviQubitMappingRouting2022}. Reduction has shown good performance in solving these problems. However, these reduction methods are designed for specific problems. They lack an automated reduction mechanism and are therefore not general. Furthermore, the existing reduction methods mainly focus on combinatorial optimization problems. This makes the reduction process easier, as MaxSAT itself is also a combinatorial optimization problem. In contrast, reducing numerical optimization problems is more challenging. Research in this area is still limited. Compared to existing reduction methods, the proposed OP-to-MaxSAT reduction automates the reduction processes for various optimization problems. It requires no manual intervention for adapting to specific problems and works for both combinatorial and numerical optimization problems.

\newpage

\subsection{Error Analysis}

The proposed OP-to-MaxSAT reduction encodes variables as signed binary fixed-point numbers. This encoding requires that all values involved in the computation can be represented as signed binary fixed-point numbers with $n$ integer bits and $m$ fractional bits $ m $. This requirement implies that OP-to-MaxSAT reduction can only guarantee the equivalence between the original and reduced problems under finite precision. Numerical errors occur when the values in the problem exceed the representable range. These errors are primarily caused by finite fractional bits. Any integer can be exactly represented by choosing a sufficiently large $n$. However, some decimal fractions, such as 0.1, cannot be exactly represented with finite binary bits. Since they have infinite binary expansions, they will always result in errors, no matter how large $m$ is.

\begin{figure}[h]
\centering
\includegraphics[width=11cm]{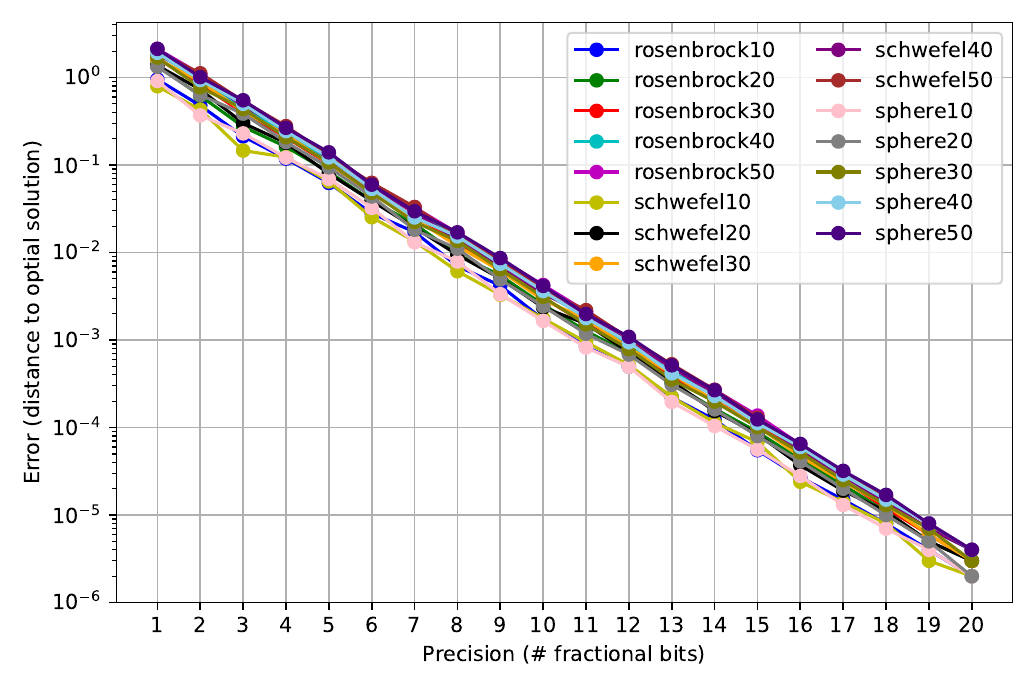}
\caption{Relationship between precision and error.}
\label{fig:error-analysis}
\end{figure}

To evaluate how precision affects numerical errors, we conducted experiments on 15 problem instances across three numerical optimization problems. The precision was defined as the number of fractional bits $m$. The error was measured as the Euclidean distance between the solution found by GORED and the theoretical optimal solution. As shown in Figure~\ref{fig:error-analysis}, the error decreases exponentially as the precision increases. All observed errors stayed within acceptable bounds. These findings demonstrate that despite precision limits, the proposed OP-to-MaxSAT reduction can maintain approximate equivalence between the original problem and the reduced problem under finite precision.

\subsection{Limitations and Future Directions}
Our method demonstrates improved generality and automation compared to existing methods. However, this generality still has limitations. For example, GORED cannot be directly applied to multi-objective optimization problems or black-box optimization problems. In the case of multi-objective optimization, GORED can only produce a single solution, whereas such problems require a Pareto solution set. A simple method to address this issue is to run GORED multiple times, each time assigning a different set of weights to the objectives to convert the problem into a single-objective formulation. One Pareto-optimal solution can be obtained from each run, and these solutions can be collected to approximate a Pareto solution set. As for black-box optimization, since OP-to-MaxSAT reduction relies on white-box constraints and objectives, GORED is not applicable unless the optimization problem is converted into a white-box formulation.

OP-to-MaxSAT reduction is the key to enabling the generality of GORED. However, the reduction may generate large MaxSAT instances, which may lead to a long solving time. Specifically, when handling complex constraints, the reduction introduces a large number of intermediate variables and hard clauses, which increases the workload of the MaxSAT solver. As a result, solving time grows significantly. This issue becomes especially serious for high-dimensional optimization problems. Consequently, GORED may not be suitable for applications with strict real-time requirements.

Future research can address the above issue from two perspectives: the reduction and the MaxSAT solver. From the perspective of reduction, more compact reduction rules can be explored to reduce the number of intermediate variables and clauses. This improvement would decrease the size of generated MaxSAT instances and improve solving efficiency. In addition, during reduction, prior knowledge about the problem structure can be passed to the MaxSAT solver for preprocessing and pruning, which may enhance performance without sacrificing generality. From the perspective of the MaxSAT solver, as technology advances, MaxSAT solvers are becoming capable of handling larger instances. Future MaxSAT solvers are expected to handle even larger problems and provide faster MaxSAT solvers for GORED, enabling both generality and efficiency. This research direction holds both theoretical and practical significance, and deserves further exploration and attention.

\section{Materials and Methods}\label{secmethod}

\subsection{Benchmark}
The problem instances used in this paper are from public benchmarks. Instances for the graph coloring problem, multi-dimensional knapsack problem, and two-echelon vehicle routing problem are obtained from OR-LIBRARY\cite{BeasleyORLibrary1990}. Traveling salesman problem instances were taken from TSPLIB\cite{gerhardTSPLIB1991}. Vehicle routing problem instances are from CVRPLIB\cite{uchoaCVRPLIB2017}. Job-shop and open-shop scheduling problem instances are from Taillard\cite{TAILLARD1993}. Quadratic assignment problem instances are obtained from QAPLIB\cite{QAPLIB1991}. The Shifted Sphere Function, Shifted Schwefel's Problem, and Shifted Rosenbrock's Function are part of the benchmarks from the IEEE CEC 2005\cite{CEC2005}.

\subsection{Selection of MaxSAT Solver}
We tested all MaxSAT solvers from the weighted complete track of MaxSAT Evaluation 2024\cite{bergMaxSATEvaluation2024} on small instances of the test problems. Based on the results, the fastest solver, Pacose24\cite{bergPacose2024}, is selected as the MaxSAT solver used in GORED for all experiments in this paper.
All experiments were conducted on a machine with an Intel(R) Core(TM) i9-10980XE CPU and 128GB of memory.

\subsection{Parameter Settings}
For the experiments on solution quality, the parameter settings of GORED for each test problem are shown in Table~\ref{tab:parameter-settings-of-GORED}. Note that the setting of these parameters only affects the size of the generated MaxSAT instances and the solving time of GORED, but has no impact on generality. We can also set the parameters to sufficiently large values (e.g., $n=20$, $m=20$) so that all problems are solved using the same parameter settings.

{
\setlength{\tabcolsep}{3pt}
\begin{table}[h]
\centering
\caption{Parameter Settings of GORED on Test Problems}
\label{tab:parameter-settings-of-GORED}
\begin{tabular}{l c c}
\toprule
Optimization Problem & \# integer bits $n$ & \# fractional bits $m$ \\
\midrule
Graph Coloring Problem              & 10  &  1  \\
Traveling Salesman Problem          & 15  &  1  \\
Capacitated Vehicle Routing Problem & 15  &  1  \\
Two-Echelon Vehicle Routing Problem & 20  &  1  \\
Multidimensional Knapsack Problem   & 15  &  5  \\
Job-Shop Scheduling Problem         & 20  &  1  \\
Open-Shop Scheduling Problem        & 20  &  1  \\
Quadratic Assignment Problem        & 10  &  1  \\
Shifted Sphere Function             & 20  & 20  \\
Shifted Schwefel's Problem          & 20  & 20  \\
Shifted Rosenbrock's Function       & 20  & 20  \\
\bottomrule
\end{tabular}
\end{table}
}

\subsection{Data Availability}

All benchmark instances is available at \url{https://github.com/YuxinZhaozyx/GORED}. All other data are included in the manuscript and its appendices.

\subsection{Code Availability}

The code of GORED is available at \url{https://github.com/YuxinZhaozyx/GORED}.

\backmatter

\section{Acknowledgments}

This work is supported by National Natural Science Foundation of China (92570116, 62276103), Innovation Team Project of General Colleges and Universities in Guangdong Province (2023KCXTD002), the Research and Development Project on Key Technologies for Intelligent Sensing and Analysis of Urban Events Based on Low-Altitude Drones (2024BQ010011), Guangdong Basic and Applied Basic Research Foundation (2023B1515120020) and National Key R\&D Program of China (2025YFC3410000).

\section{Author Contributions}

Y.Z, H.H and Z.H designed research; Y.Z and H.H performed research; Y.Z wrote the paper with contributions from all co-authors.

\section{Competing Interests}

The authors declare no competing interest.


\bibliography{sn-bibliography}

\newgeometry{top=2.1cm, bottom=2.1cm, left=1.6cm, right=1.6cm}
\appendixpage
\appendix
\section{Grammar of Unified Modeling Language}\label{appendix:grammar}
The grammar of the proposed modeling language is described below using Extended Backus-Naur Form (ISO/IEC 14977).

\begin{lstlisting}
model = {'\begin{align}', (objective|constraint), {'\\', (objective|constraint)}, ['\\'], '\end{align}'};

objective = ('\max' | '\min'), ['&&'], numeric_expression, ['&&' | '&&', conditions];
constraint = ['s.t.'], ['&&'], relation_expression, ['&&' | '&&', conditions];

conditions = condition, {',', condition};
condition = variable, {',', variable}, '=', numeric_expression, ',', ('...' | '\dots' | '\cdots'), ',',           numeric_expression
          | ['\forall'], variable, {',', variable}, '\in', set_expression
          | ['\forall'], variable, {',', variable}, ('\subset'|'\subseteq'|'\subsetneq'), set_expression
          | numeric_expression, ('\le' | '<'), variable, {('\le' | '<'), variable}, ('\le' | '<'), numeric_expression
          | relation_expression;

relation_expression = variable, '=', variable, {'=', variable}
                    | variable, '\neq', variable
                    | numeric_expression, '=', numeric_expression, {'=', numeric_expression}
                    | numeric_expression, '\neq', numeric_expression
                    | numeric_expression,('\le'|'<'),numeric_expression,{('\le'|'<'),numeric_expression}
                    | numeric_expression,('\ge'|'>'),numeric_expression,{('\ge'|'>'),numeric_expression}
                    | set_expression, '=', set_expression, {'=', set_expression}
                    | set_expression, '\neq', set_expression
                    | set_expression, ('\subset' | '\subseteq' | '\subsetneq'), set_expression
                    | numeric_expression, ('\in' | '\notin' | '\not\in'), set_expression;

numeric_expression = ['-'], numeric_1st_term, {('+' | '-'), numeric_1st_term};

numeric_1st_term = numeric_2nd_term {['\cdot' | '\times' | '/'], numeric_2nd_term};

numeric_2nd_term = numeric_3rd_term
                 | '\frac', '{', numeric_expression, '}', ('{', numeric_expression, '}' | one_token)
                 | numeric_3rd_term, '^', ('{', numeric_expression, '}' | one_token)
                 | ('\sum' | '\prod'), '_', '{', variable, '=', numeric_expression, [',', conditions], '}', '^', ('{', numeric_expression, '}' | one_token) numeric_1st_term
                 | ('\sum' | '\prod'), '_', '{', conditions, '}', numeric_1st_term
                 | ('\max' | '\min'), set_expression
                 | ('\max' | '\min'), '_', '{', variable, '=', numeric_expression, [',', conditions], '}', '^', ('{', numeric_expression, '}' | one_token), ('\{', numeric_expression, '\}' | '\set', '{', numeric_expression, '}')
                 | ('\max' | '\min'), '_', '{', conditions, '}', ('\{', numeric_expression, '\}' | '\set', '{', numeric_expression, '}');

numeric_3rd_term = variable
                 | numeric
                 | '(', numeric_expression, ')'
                 | ('|' | '\vert'), (variable | numeric_expression | set_expression), ('|' | '\vert')
                 | '\mathbb{I}', '(', relation_expression, ')'
                 | '\lfloor', numeric_expression, '\rfloor'
                 | '\lceil', numeric_expression, '\rceil';

set_expression = set_term
               | set_expression, ('\cup' | '\cap' | '\setminus' | '\backslash'), set_expression;

set_term = variable
         | set
         | '(', set_expression, ')'
         | ('\bigcup' | '\bigcap'), '_', '{', variable, '=', numeric_expression, [',', conditions], '}', '^', ('{', numeric_expression, '}' | one_token) set_term
         | ('\bigcup' | '\bigcap'), '_', '{', conditions, '}', set_term;

numeric = MULTI_TOKEN_NUMERIC
        | DIGIT;

set = predefined_set
    | '\set', '{', numeric_expression, {',', numeric_expression}, '}'
    | '\{', numeric_expression, {',', numeric_expression}, '\}'
    | '\set', '{', numeric_expression, ',', ('\dots' | '...' | '\cdots'), ',', numeric_expression, '}'
    | '\{', numeric_expression, ',', ('\dots' | '...' | 'cdots'), ',', numeric_expression, '\}';

predefined_set = '\mathbb{R}' | '\mathbb{R}^+' | '\mathbb{R}^-' | '\mathbb{Z}' | '\mathbb{Z}^+'
               | '\mathbb{Z}^-' | '\mathbb{N}' | '\emptyset';

variable = VARIABLE_TERM
         | VARIABLE_TERM, '_', ('{', numeric_expression, {',', numeric_expression}, '}' | one_token);

one_token = DIGIT | VARIABLE_TERM;
\end{lstlisting}

\newpage

\begin{lstlisting}[firstnumber=70]
MULTI_TOKEN_NUMERIC = ('0' | DIGIT_EXCEPT_ZERO, {DIGIT}), ['.', {DIGIT}]; 

DIGIT = '0' | DIGIT_EXCEPT_ZERO;

DIGIT_EXCEPT_ZERO = '1' | '2' | '3' | '4' | '5' | '6' | '7' | '8' | '9';

VARIABLE_TERM = LETTER
              | GREEK_LETTER
              | '\mathcal', '{', LETTER, '}'
              | '\mathbf', '{' , LETTER, '}'
              | '\boldsymbol', '{', (LETTER | GREEK_LETTER), '}';

LETTER = 'a' | 'b' | 'c' | 'd' | 'e' | 'f' | 'g' | 'h' | 'i' | 'j' | 'k' | 'l' | 'm' | 'n' | 'o' | 'p'
       | 'q' | 'r' | 's' | 't' | 'u' | 'v' | 'w' | 'x' | 'y' | 'z' | 'A' | 'B' | 'C' | 'D' | 'E' | 'F' 
       | 'G' | 'H' | 'I' | 'J' | 'K' | 'L' | 'M' | 'N' | 'O' | 'P' | 'Q' | 'R' | 'S' | 'T' | 'U' | 'V' 
       | 'W' | 'X' | 'Y' | 'Z';

GREEK_LETTER = '\Alpha' | '\Beta' | '\Gamma' | '\Delta' | '\Epsilon' | '\Zeta' | '\Eta' | '\Theta'
             | '\Iota' | '\Kappa' | '\Lambda' | '\Mu' | '\Nu' | '\Xi' | '\Omicron' | '\Pi' | '\Rho'
             | '\Sigma' | '\Tau' | '\Upsilon' | '\Phi' | '\Chi' | '\Psi' | '\Omega' | '\alpha' | '\beta'
             | '\gamma' | '\delta' | '\epsilon' | '\zeta' | '\eta' | '\theta' | '\iota' | '\kappa' 
             | '\lambda' | '\mu' | '\nu' | '\xi' | '\omicron' | '\pi' | '\rho' | '\sigma' | '\tau'
             | '\upsilon' | '\phi' | '\chi' | '\psi' | '\omega' | '\varepsilon' | '\vartheta'
             | '\varkappa' | '\varpi' | '\varrho' | '\varsigma' | '\varphi';
\end{lstlisting}

\section{Mathematical Models of the Test Problems}\label{appendix:models}
The mathematical models of the test problems used in the experiments are represented in the proposed modeling language as follows:

\subsection{Graph Coloring Problem} The Graph Coloring Problem\cite{BeasleyORLibrary1990} involves assigning colors to the vertices of a given undirected connected graph such that each vertex receives exactly one color, and no two adjacent vertices share the same color. The objective is to determine a valid coloring scheme that minimizes the total number of distinct colors used.
The mathematical model of this problem represented in the proposed modeling language is shown in Formulas~\ref{formula:gcp-start}-\ref{formula:gcp-end}.
\begin{align}
    \min && & \max_{i=1}^n \{x_i\} \label{formula:gcp-start} \\
    && & x_i \neq x_j && i,j = 1,\dots,n, \space i < j, \space c_{i,j} = 1 \\
    && & x_i \in \{1,\dots,n\} && i = 1,\dots,n \label{formula:gcp-end}
\end{align}
where $x_i$ denotes the color assigned to vertex $i$, $c_{i,j}$ indicates whether vertices $i$ and $j$ are adjacent.

\subsection{Traveling Salesman Problem}
The Traveling Salesman Problem\cite{gerhardTSPLIB1991} is defined as follows: Given a set of $n$ cities, a salesman must start from an initial city, visit each of the remaining cities exactly once, and return to the starting city. The objective is to find the shortest path that satisfies these constraints.
The mathematical model of this problem represented in the proposed modeling language is shown in Formulas~\ref{formula:tsp-start}-\ref{formula:tsp-end}.
\begin{align}
\min && & \sum_{i=1}^n \sum_{j=1,j\neq i}^n c_{i,j} x_{i,j} \label{formula:tsp-start}  \\
s.t. && & x_{i,j} \in \{0,1\} && i,j = 1,\dots,n \\
&& & \sum_{i=1, i\neq j}^n x_{i,j} = 1 && j = 1,\dots,n \\
&& & \sum_{j=1, j\neq i}^n x_{i,j} = 1 && i = 1,\dots,n \\
&& & u_i - u_j + n x_{i,j} \le n-1 && 2 \le i \le n, \space  2\le j\le n, \space i\neq j \\
&& & u_i \in \{2, \dots, n\} && i = 2,\dots,n \label{formula:tsp-end}
\end{align}
where $n$ is the number of cites, $x_{i,j}$ indicates whether the salesman travels from city $i$ to city $j$, $c_{i,j}$ denotes the distances between city $i$, and city $j$, and $u_i$ represents the order in which city $i$ is visited by the salesman.

\subsection{Quadratic Assignment Problem}
The Quadratic Assignment Problem\cite{QAPLIB1991} is defined as follows: Given $n$ facilities and $n$ distinct locations, the objective is to assign each facility to a unique location such that the total cost is minimized. The total cost is determined by the sum of the products of the flow between each pair of facilities and the distance between their assigned locations, weighted by the associated transportation cost.
The mathematical model of this problem represented in the proposed modeling language is shown in Formulas~\ref{formula:qap-start}-\ref{formula:qap-end}.
\begin{align}
\min && & \sum_{i=1}^n \sum_{j=1}^n \sum_{k=1}^n \sum_{l=1}^n f_{i,j} d_{k,l} x_{i,k} x_{j,l} \label{formula:qap-start} \\
s.t. && & x_{i,j}\in \{0,1\} && i,j=1,\dots,n  \\
&& & \sum_{i=1}^n x_{i,j} = 1 && j = 1, \dots, n \\
&& & \sum_{j=1}^n x_{i,j} = 1 && i = 1, \dots, n \label{formula:qap-end}
\end{align}
where $x_{i,j}$ indicates whether facility $i$ is assigned to location $j$, $f_{i,j}$ denotes the transportation cost from facility $i$ to facility $j$, and $d_{i,j}$ denotes the distance from facility $i$ to facility $j$.

\subsection{Capacitated Vehicle Routing Problem}
The Capacitated Vehicle Routing Problem\cite{uchoaCVRPLIB2017} is defined as follows: Given a depot, $n-1$ customer locations, and a fleet of $n$ vehicles, each with a capacity of $Q$, the objective is to determine a set of routes for the vehicles such that: (1) Each route starts and ends at the depot; (2) Each customer is visited exactly once by exactly one vehicle; (3) The total demand of the customers served on any single route does not exceed the vehicle capacity $Q$; (4) The required quantity of goods is delivered to each customer upon visit. The goal is to minimize the total travel distance across all vehicles.
The mathematical model of this problem represented in the proposed modeling language is shown in Formulas~\ref{formula:cvrp-start}-\ref{formula:cvrp-end}.
\begin{align}
\min && & \sum_{k=1}^m \sum_{i=1}^n \sum_{j=1}^n d_{i,j} x_{i,j,k} \label{formula:cvrp-start} \\
s.t. && & x_{i,j,k} \in \{0,1\} && \forall k \in \{1, \dots, m\}, i,j \in \{1,\dots,n\} \\
&& & x_{i,i,k} = 0 && \forall k \in \{1, \dots, m\}, i\in \{1,\dots,n\} \\
&& & \sum_{i=1}^n x_{i,j,k} = \sum_{i=1}^n x_{j,i,k} && \forall j \in \{1, \dots,n\},k\in \{1,\dots,m\} \\
&& & \sum_{k=1}^m \sum_{i=1}^n x_{i,j,k} = 1 && \forall j \in \{2, \dots, n\} \\
&& & \sum_{j=2}^n x_{1,j,k} = 1 && \forall k \in \{1, \dots, m\} \\
&& & \sum_{i=1}^n \sum_{j=2}^n q_j x_{i,j,k} \le Q && \forall k \in \{1,\dots,m\} \\
&& & u_j - u_i - q_j + Q \ge  Q \max_{k=1}^m \{ x_{i,j,k} \} && \forall i,j \in \{2,\dots,n\}, \space i\neq j \\
&& & u_i \in \{q_i, \dots, Q\} && \forall i \in \{2,\dots,n\} \label{formula:cvrp-end}
\end{align}
where $n$ is the number of locations, the first location is the depot, the rest are locations of customers, $m$ is the number of vehicles, $x_{i,j,k}$ indicates whether vehicle $k$ travels from location $i$ to location $j$, $d_{i,j}$ denotes the distance from location $i$ to location $j$, $q_i$ denotes the demand of customer $i$, $Q$ is the capacity of each vehicle, and $u_i$ denotes the cumulative amount of goods delivered by a vehicle upon completing service at location $i$.

\subsection{Two-Echelon Vehicle Routing Problem}
The Two-Echelon Vehicle Routing Problem\cite{BeasleyORLibrary1990} is defined on a two-level distribution system comprising one central depot, $m$ satellites, $n$ customers, $N_1$ first-level vehicles, and $N_2$ second-level vehicles. 
All customer demands are known and must be satisfied. Goods are initially located at the depot and must be transferred to customers via the satellites: first-level vehicles move goods from the depot to the satellites, and second-level vehicles then distribute them from the satellites to the customers. Each customer is visited exactly once by a single second-level vehicle.
The objective is to determine the number of vehicles used at each level and to design routes for first-level and second-level vehicles such that all customer demands are met and the total travel distance across all vehicles is minimized.
The mathematical model of this problem represented in the proposed modeling language is shown in Formulas~\ref{formula:2evrp-start}-\ref{formula:2evrp-end}.
\begin{align}
\min && & \sum_{i, j \in \{0\} \cup S, i \neq j } c_{i,j} y_{i,j} + \sum_{i \in S} \sum_{j \in C} ( c_{i, j} z_{i,j,i} + c_{j, i} z_{j,i,i} ) + \sum_{k \in S} \sum_{i, j \in C, i \neq j} c_{i,j} z_{i,j,k} \qquad \qquad \qquad \label{formula:2evrp-start}
\end{align}
\begin{align}
s.t. && & \sum_{i \in S} y_{0, i} \le N_1 \\
&& & \sum_{i \in \{0\} \cup S, i \neq j} y_{i, j} = \sum_{i\in \{0\} \cup S, i \neq j} y_{j, i} && \forall j \in S \\
&& & \sum_{i \in S} \sum_{j \in C} z_{i,j, i} \le N_2 \\
&& & \sum_{i \in S \cup C, i \neq j} z_{i,j,k} = \sum_{i \in S \cup C, i \neq j} z_{j,i,k} && \forall k \in S, j \in C \\
&& & z_{i,j,k} = 0 && \forall k \in S, i,j \in S, i \neq j \\
&& & z_{i,j,k} = 0 && \forall k \in S, i \in S, j \in C, i \neq k \\
&& & z_{j,i,k} = 0 && \forall k \in S, i \in S, j \in C, i \neq k \\
&& & Q_{1,i,j} \le y_{i,j} T_1 && \forall i,j \in \{0\} \cup S, i \neq j \\
&& & Q_{2,i,j} \le \sum_{k \in S} z_{i,j,k} T_2 && \forall i, j \in S \cup C, i \neq j \\
&& & Q_{1,i,0} = 0 && \forall i \in S \\
&& & Q_{2,i,j} = 0 && \forall i \in S \cup C, j \in S, i \neq j \\
&& & \sum_{i \in \{0\} \cup S, i \neq j} Q_{1,i,j} = \sum_{k \in \{0\} \cup S, k \neq j} Q_{1,j,k} + \sum_{l \in C} Q_{2,j,l} && \forall j \in S \\
&& & \sum_{i \in S \cup C, i \neq j} Q_{2,i,j} = \sum_{k \in S \cup C, k \neq j} Q_{2,j,k} + \rho_j && \forall j \in C \\
&& & \sum_{k \in S} \sum_{i \in S \cup C, i\neq j} z_{i, j, k} = 1 && \forall j \in C \\
&& & z_{i, j, k} \in \{0, 1\} && \forall k \in S, i,j \in S \cup C, i \neq j \\
&& & y_{i,j} \in \{0, ..., N_1\} && \forall i, j \in \{0\} \cup S, i \neq j \\
&& & S = \{1, \dots, m\} \\
&& & C = \{m + 1, \dots, m + n\} \\
&& & Q_{1,i,j} \in \{0, \dots, N_1 T_1\} && \forall i,j \in \{0\} \cup S, i \neq j \\
&& & Q_{2,i,j} \in \{0, \dots, T_2\} && \forall i,j \in S \cup C, i \neq j \label{formula:2evrp-end}
\end{align}
where there are $m+n+1$ positions, position $0$ is the depot, positions $1$ to $m$ are satellites, denoted by set $S$, positions $m+1$ to $m+n$ are customers, denoted by set $C$. $N_1$ and $N_2$ are the number of candidate first-level and second-level vehicles, respectively. $c_{i,j}$ denotes the distance from position $i$ to position $j$. $y_{i,j}$ denotes the number of first-level vehicles that travel from position $i$ to position $j$. $z_{i,j,k}$ indicates whether second-vehicle $k$ travels from position $i$ to position $j$. $\rho_j$ is the demand of customer $j$. $Q_{k,i,j}$ denotes the amount of goods carried by a vehicle at level $k$ on the route segment from location $i$ to location $j$. $T_k$ represents the capacity of vehicles at level $k$.

\subsection{Multidimensional Knapsack Problem}
The Multidimensional Knapsack Problem\cite{BeasleyORLibrary1990} is defined as follows: Given a set of $n$ items and a single knapsack, each item has an associated value and $m$ distinct weight attributes. The knapsack has a capacity constraint on each of the $m$ weight dimensions. The objective is to select a subset of items to place into the knapsack such that the total value of the selected items is maximized, while ensuring that the sum of weights in each dimension does not exceed the corresponding capacity constraint of the knapsack.
The mathematical model of this problem represented in the proposed modeling language is shown in Formulas~\ref{formula:mkp-start}-\ref{formula:mkp-end}.
\begin{align}
\max && & \sum_{i=1}^n v_i x_i \label{formula:mkp-start} \\
s.t. && & \sum_{i=1}^n w_{i,j} x_i \le W_j && j=1,\dots,m \\
&& & x_i \in \{0, 1\} && i=1,\dots,n \label{formula:mkp-end}
\end{align}
where $n$ is the number of items, $m$ is the number of weight dimensions, $v_i$ is the value of item $i$, $w_{i,j}$ is the weight of item $i$ on dimension $j$, $W_j$ is the capacity of the $j$-th weight dimension, and $x_i$ indicates whether item $i$ is selected to place into the knapsack.

\subsection{Job-Shop Scheduling Problem}
The Job-Shop Scheduling Problem\cite{TAILLARD1993} is defined as follows: Given $n$ jobs and $m$ machines, each job must be processed on each machine exactly once. The processing order and time across machines are predetermined and may differ for each job. The constraints are that (1) each machine can process at most one job at a time, and (2) each job can be processed on only one machine at a time. The objective is to determine a feasible schedule that minimizes the makespan, i.e., the total time required to complete all jobs.
The mathematical model of this problem represented in the proposed modeling language is shown in Formulas~\ref{formula:jsp-start}-\ref{formula:jsp-end}.
\begin{align}
\min && & C && \label{formula:jsp-start} \\
s.t. && & 0 \le x_{i,j} \le V && \forall j\in J,i\in M \\
&& & x_{\sigma_{j, h} ,j} \ge x_{\sigma_{j, h-1} ,j} + p_{\sigma_{j,h-1},j} && \forall j\in J, h=2,\dots,m \\
&& & x_{i,j} \ge x_{i,k} + p_{i,k} - 2V \cdot z_{i,j,k} && \forall j,k \in J, j<k, i\in M \\
&& & x_{i,k} \ge x_{i,j} + p_{i,j} - 2V \cdot (1 - z_{i,j,k}) && \forall j,k \in J, j<k,i\in M \\
&& & x_{\sigma_{j,m},j} + p_{\sigma_{j,m},j} \le C && \forall j\in J \\
&& & z_{i,j,k} \in \{0,1\} && \forall j,k \in J, j<k, i \in M \\
&& & V= \sum_{j\in J}\sum_{i \in M} p_{i,j} \\
&& & J=\{1,\dots, n\} \\
&& & M=\{1,\dots, m\} \\
&& & x_{i,j} \in \mathbb{Z} && \forall j \in J, i \in M \label{formula:jsp-end}
\end{align}
where $C$ denotes the makespan, $J$ is the set of jobs with $|J| = n$, and $M$ is the set of machines with $|M| = m$. $\sigma_{j,h}$ represents the machine required for the $h$-th operation of job $j$, where each job consists of exactly $m$ operations, and no two operations of the same job are processed on the same machine. $x_{i,j}$ denotes the start time of job $j$ on machine $i$, and $p_{i,j}$ is the processing time of job $j$ on machine $i$. The binary variable $z_{i,j,k}$ indicates whether job $j$ starts before job $k$ on machine $i$. $V$ is a sufficiently large positive constant (big-M value).

\subsection{Open-Shop Scheduling Problem}
The Open Shop Scheduling Problem\cite{TAILLARD1993} is defined as follows: Given $n$ jobs and $m$ machines, each job must be processed exactly once on each machine, but the processing order of machines for any job is not predetermined and can be chosen freely. The processing time of each job on each machine is given and may vary across jobs and machines. The constraints are that (1) each machine can process at most one job at a time, and (2) each job can be processed on only one machine at a time. The objective is to determine a feasible schedule, including the processing sequence on each machine and for each job, that minimizes the makespan, i.e., the total time required to complete all jobs.
The mathematical model of this problem represented in the proposed modeling language is shown in Formulas~\ref{formula:osp-start}-\ref{formula:osp-end}.
\begin{align}
\min && & C \label{formula:osp-start} \\
s.t. && & 0 \le x_{i,j} \le V && \forall j\in J,i\in M \\
&& & x_{k,i} \ge x_{j,i} + p_{j,i} - 2V \cdot w_{i,j,k} && \forall i \in J, j,k \in M, j < k \\
&& & x_{j,i} \ge x_{k,i} + p_{k,i} - 2V \cdot (1 - w_{i,j,k}) && \forall i \in J, j,k \in M, j < k \\
&& & x_{i,j} \ge x_{i,k} + p_{i,k} - 2V \cdot z_{i,j,k} && \forall i\in M, j,k\in J, j < k \\
&& & x_{i,k} \ge x_{i,j} + p_{i,j} - 2V \cdot (1 - z_{i,j,k}) && \forall i\in M, j,k\in J, j < k \\
&& & x_{i,j} + p_{i,j} \le C && \forall i\in M, j\in J \\
&& & z_{i,j,k} \in \{0,1\} && \forall i\in M, j,k\in J, j < k \\
&& & w_{i,j,k} \in \{0,1\} && \forall i\in J, j,k\in M, j < k \\
&& & V = \sum_{i\in M}\sum_{j\in J} p_{i,j} \\
&& & J=\{1,\dots, n\} \\
&& & M=\{1,\dots, m\} \\
&& & x_{i,j} \in \mathbb{Z} && \forall j \in J, i \in M \label{formula:osp-end}
\end{align}
where $C$ denotes the makespan, $J$ is the set of jobs with $|J| = n$, and $M$ is the set of machines with $|M| = m$. $x_{i,j}$ denotes the start time of job $j$ on machine $i$, and $p_{i,j}$ is the processing time of job $j$ on machine $i$. The binary variable $z_{i,j,k}$ indicates whether job $j$ starts before job $k$ on machine $i$. The binary variable $w_{i,j,k}$ indicates whether job $i$ starts on machine $j$ before it starts on machine $k$. $V$ is a sufficiently large positive constant (big-M value).

\subsection{Shifted Sphere Function}
The Shifted Sphere Function is a benchmark from IEEE CEC 2005\cite{CEC2005}.
The mathematical model of this problem represented in the proposed modeling language is shown in Formulas~\ref{formula:sphere-start}-\ref{formula:sphere-end}.
\begin{align}
\min && & \sum_{i=1}^D z_i^2 + b \label{formula:sphere-start} \\
s.t. && & z_i = x_i - o_i && i=1,\dots,D \\
     && & b = -450 \\
     && & -100 \le x_i \le 100 && i=1,\dots,D \label{formula:sphere-end}
\end{align}
where $x_1$ to $x_D$ are the decision variables, the sequence $(o_1,o_2,\dots,o_D)$ denotes the optimal solution, and $b$ is the optimal objective value.

\subsection{Shifted Schwefel's Problem}
The Shifted Schwefel's Problem is a benchmark from IEEE CEC 2005\cite{CEC2005}.
The mathematical model of this problem represented in the proposed modeling language is shown in Formulas~\ref{formula:schwefel-start}-\ref{formula:schwefel-end}.
\begin{align}
\min && & \sum_{i=1}^D (\sum_{j=1}^i z_j)^2 + b \label{formula:schwefel-start} \\
s.t. && & z_i = x_i - o_i && i=1,\dots,D \\
     && & b = -450 \\
     && & -100 \le x_i \le 100 && i=1,\dots,D \label{formula:schwefel-end}
\end{align}
where $x_1$ to $x_D$ are the decision variables, the sequence $(o_1,o_2,\dots,o_D)$ denotes the optimal solution, and $b$ is the optimal objective value.

\subsection{Shifted Rosenbrock's Function}
The Shifted Rosenbrock's Function is a benchmark from IEEE CEC 2005\cite{CEC2005}.
The mathematical model of this problem represented in the proposed modeling language is shown in Formulas~\ref{formula:rosenbrock-start}-\ref{formula:rosenbrock-end}.
\begin{align}
\min && & \sum_{i=1}^{D-1} (100 (z_i^2 - z_{i+1})^2 + (z_i - 1)^2) + b 
\label{formula:rosenbrock-start} \\
s.t. && & z_i = x_i - o_i + 1 && i=1,\dots,D \\
     && & b = -390 \\
     && & -100 \le x_i \le 100 && i=1,\dots,D \label{formula:rosenbrock-end}
\end{align}
where $x_1$ to $x_D$ are the decision variables, the sequence $(o_1,o_2,\dots,o_D)$ denotes the optimal solution, and $b$ is the optimal objective value.

\section{Reduction Rules of OP-to-MaxSAT Reduction}\label{appendix:reduction-rules}
In the OP-to-MaxSAT reduction, numerical variables are encoded as signed binary fixed-point numbers, shown in Formula~\ref{formula:encoding-variable}.
The reduction rules for reducing constraints and the objective of optimization problems are as follows:

\subsection{Reduction Rule for the Full Adder: $FullAdder(x,y,c_{in},z,c_{out})$} 
The rule models a full adder that takes Boolean inputs $x$, $y$, and $c_{in}$, and produces the sum $z$ and the carry-out $c_{out}$. This rule is proposed in \cite{zhouOptimizingSATEncodings2017}. The clauses generated by this rule are shown in Formula~\ref{eq:full-adder-rule}.
\begin{equation}
\begin{array}{c}
 x \lor \neg y \lor c_{in} \lor z \\
 x \lor y \lor \neg c_{in} \lor z \\
 \neg x \lor \neg y \lor c_{in} \lor \neg z \\
 \neg x \lor y \lor \neg c_{in} \lor \neg z \\
 \neg x \lor c_{out} \lor z \\
 x \lor \neg c_{out} \lor \neg z \\
 \neg y \lor \neg c_{in} \lor c_{out} \\
 y \lor c_{in} \lor \neg c_{out} \\
 \neg x \lor \neg y \lor \neg c_{in} \lor z \\
 x \lor y \lor c_{in} \lor \neg z \\
\end{array}
\label{eq:full-adder-rule}
\end{equation}

This rule generates 0 intermediate Boolean variables and 10 clauses.

\subsection{Reduction Rule for the Half Adder: $HalfAdder(x,y,z,c_{out})$}
The rule models a half adder that takes Boolean inputs $x$ and $y$, and produces the sum $z$ and the carry-out $c_{out}$. The clauses generated by this rule are shown in Formula~\ref{eq:half-adder-rule}.
\begin{equation}
\begin{array}{c}
 x \lor \neg y \lor z \\
 \neg x \lor \neg y \lor \neg z \\
 \neg x \lor c_{out} \lor z \\
 x \lor \neg c_{out} \lor \neg z \\
 y \lor \neg c_{out} \\
 x \lor y \lor \neg z \\
\end{array}
\label{eq:half-adder-rule}
\end{equation}

This rule generates 0 intermediate Boolean variables and 6 clauses.

\subsection{Reduction Rule for the Two's Complement: $Complement(\boldsymbol a, \boldsymbol a')$}
The rule models the relation between a signed binary fixed-point number $\boldsymbol a = a_{m+n} a_{m+n-1}\cdots a_0$ and its two's complement encoding $\boldsymbol a' = a'_{m+n} a'_{m+n-1}\cdots a'_0$. The clauses generated by this rule are shown in Formula~\ref{eq:complement-rule}.
\begin{equation}
\begin{array}{cl}
a_{m+n} \lor \neg a_i \lor a'_i & i=0,\dots,m+n-1 \\
a_{m+n} \lor a_i \lor \neg a'_i & i=0,\dots,m+n-1 \\
a_{m+n} \lor \neg a'_{m+n} \\
\neg a_{m+n} \lor d_0 \\
\neg a_{m+n} \lor HalfAdder(\neg a_i, d_i, a'_i, d_{i+1}) & i=0,\dots,m+n-1 \\
\neg a_{m+n} \lor d_{m+n} \lor a'_{m+n} \\
\neg a_{m+n} \lor \neg d_{m+n} \lor \neg a'_{m+n} \\
\bigvee_{i=0}^{m+n-1} a'_i \lor \neg a'_{m+n} \\
\end{array}
\label{eq:complement-rule}
\end{equation}
where $d_0$ to $d_{m+n}$ are intermediate Boolean variables.

This rule generates $m+n+1$ intermediate Boolean variables and $8m + 8n + 5$ clauses.

\subsection{Reduction Rule for the Adder Supporting Two Inputs: $Adder(\boldsymbol a',\boldsymbol b',\boldsymbol c')$}
The rule models an adder that takes the two's complement encodings $\boldsymbol a'$ and $\boldsymbol b'$ as inputs, and produces the two's complement of the sum $\boldsymbol c'$. The clauses generated by this rule are shown in Formula~\ref{eq:adder-rule}.
\begin{equation}
\begin{array}{cl}
\neg d_0 \\
\neg a'_{m+n} \lor \neg b'_{m+n} \lor c'_{m+n} \\
a'_{m+n} \lor b'_{m+n} \lor \neg c'_{m+n} \\
a'_i \lor b'_i \lor \neg d_{i+1} & i=0,\dots,m+n \\
\neg a'_i \lor \neg b'_i \lor d_{i+1} & i=0,\dots,m+n \\
FullAdder(a'_i, b'_i, d_i, c'_i, d_{i+1}) & i=0,\dots,m+n \\
\end{array}
\label{eq:adder-rule}
\end{equation}
where $d_0$ to $d_{m+n}$ are intermediate Boolean variables.

This rule generates $m+n+1$ intermediate Boolean variables and $12m + 12n + 15$ clauses.

\subsection{Reduction Rule for the Adder Supporting Multiple Inputs: $MultiAdder(\{\boldsymbol a'_1, \boldsymbol a'_2, \cdots, \boldsymbol a'_k\}, \boldsymbol c')$}
The rule models a more powerful adder that takes the two's complement encodings $\boldsymbol a'_1$ to $\boldsymbol a'_k$ as inputs ($k \ge 2$), and produces the two's complement of the sum $\boldsymbol c'$. The clauses generated by this rule are shown in Formula~\ref{eq:multi-adder-rule}.
\begin{equation}
\begin{array}{ll}
(k = 2) \to Adder(\boldsymbol a'_1, \boldsymbol a'_2, \boldsymbol c') \\
(k = 3) \to 
\begin{Bmatrix}
Adder(\boldsymbol a'_1, \boldsymbol a'_2, \boldsymbol e') \\
Adder(\boldsymbol e', \boldsymbol a'_3, \boldsymbol c')
\end{Bmatrix} \\
(k \ge 4) \to
\begin{Bmatrix}
MultiAdder(\{\boldsymbol a'_1, \cdots, \boldsymbol a'_{\lfloor k / 2 \rfloor}\}, \boldsymbol b'_1) \\
MultiAdder(\{\boldsymbol a'_{\lfloor k / 2 \rfloor + 1}, \cdots, \boldsymbol a'_k\}, \boldsymbol b'_2) \\
Adder(\boldsymbol b'_1, \boldsymbol b'_2, \boldsymbol c') \\
\end{Bmatrix} \\
\end{array}
\label{eq:multi-adder-rule}
\end{equation}
where $\boldsymbol e'$, $\boldsymbol b'_1$, and $\boldsymbol b'_2$ are intermediate variables.

This rule generates $(2k-3)m + (2k-3)n + (2k-3)$ intermediate Boolean variables and $12(k-1)m + 12(k-1)n + 15(k-1)$ clauses.

\subsection{Reduction Rule for Sum Expression: $Sum(\{\boldsymbol a_1, \boldsymbol a_2, \cdots, \boldsymbol a_k\}, \boldsymbol c)$}
The rule models the expression $\boldsymbol c = \sum_{i=1}^k \boldsymbol a_i$, where $\boldsymbol a_i$ and $\boldsymbol c$ are signed binary fixed-point numbers. The clauses generated by this rule are shown in Formula~\ref{eq:sum-rule}.
\begin{equation}
\begin{array}{cl}
Complement(\boldsymbol a_i, \boldsymbol a'_i) & i=1,\dots,k \\
MultiAdder(\{\boldsymbol a'_1, \cdots, \boldsymbol a'_k\}, \boldsymbol c') \\
Complement(\boldsymbol c, \boldsymbol c')
\end{array}
\label{eq:sum-rule}
\end{equation}

This rule generates $(4k-1)m + (4k-1)n + (4k-1)$ intermediate Boolean variables and $(20k - 4)m +(20k-4)n + (20k-10)$ clauses.

\subsection{Reduction Rule for Equal Expression: $Equal(\boldsymbol a, \boldsymbol b)$}
The rule models the expression $\boldsymbol a = \boldsymbol b$, where $\boldsymbol a$ and $\boldsymbol b$ are signed binary fixed-point numbers. The clauses generated by this rule are shown in Formula~\ref{eq:equal-rule}.
\begin{equation}
\begin{array}{cl}
\neg a_{m+n} \lor \neg b_{m+n} \lor e \\
a_{m+n} \lor b_{m+n} \lor e \\
a_{m+n} \lor \neg b_{m+n} \lor \neg e \\
\neg a_{m+n} \lor b_{m+n} \lor \neg e \\
\neg e \lor \neg a_i \lor b_i & i=0,\dots,m+n-1 \\
\neg e \lor a_i \lor \neg b_i & i=0,\dots,m+n-1 \\
e \lor \neg a_i & i=0,\dots,m+n-1 \\
e \lor \neg b_i & i=0,\dots,m+n-1 \\
\end{array}
\label{eq:equal-rule}
\end{equation}
where $e$ is an intermediate Boolean variable.

This rule generates 1 intermediate Boolean variable and $4m+4n+4$ clauses.

\subsection{Reduction Rule for Not Equal Expression: $NotEqual(\boldsymbol a, \boldsymbol b)$}
The rule models the expression $\boldsymbol a \neq \boldsymbol b$, where $\boldsymbol a$ and $\boldsymbol b$ are signed binary fixed-point numbers. The clauses generated by this rule are shown in Formula~\ref{eq:not-equal-rule}.
\begin{equation}
\begin{array}{cl}
\neg a_{m+n} \lor \neg b_{m+n} \lor e \\
a_{m+n} \lor b_{m+n} \lor e \\
a_{m+n} \lor \neg b_{m+n} \lor \neg e \\
\neg a_{m+n} \lor b_{m+n} \lor \neg e \\

\neg e \lor \neg a_i \lor b_i \lor p_i & i=0,\dots,m+n-1 \\
\neg e \lor a_i \lor \neg b_i \lor p_i & i=0,\dots,m+n-1 \\
\neg e \lor \neg a_i \lor \neg b_i \lor \neg p_i & i=0,\dots,m+n-1 \\
\neg e \lor a_i \lor b_i \lor \neg p_i & i=0,\dots,m+n-1 \\
\neg e \lor \bigvee_{i=0}^{m+n-1} p_i \\
e \lor \bigvee_{i=0}^{m+n-1} a_i \vee \bigvee_{i=0}^{m+n-1} b_i \\
\end{array}
\label{eq:not-equal-rule}
\end{equation}
where $e$ and $p_i$ are intermediate Boolean variables.

This rule generates $m+n+1$ intermediate Boolean variables and $4m+4n+6$ clauses.

\subsection{Reduction Rule for Less-Than Expression: $LessThan(\boldsymbol a, \boldsymbol b)$}
The rule models the expression $\boldsymbol a < \boldsymbol b$, where $\boldsymbol a$ and $\boldsymbol b$ are signed binary fixed-point numbers. The clauses generated by this rule are shown in Formula~\ref{eq:less-than-rule}.
\begin{equation}
\begin{array}{cl}
a_{m+n} \lor \neg b_{m+n} \\
\neg a_{m+n} \lor b_{m+n} \lor \bigvee_{i=0}^{m+n-1} a_i \lor \bigvee_{i=0}^{m+n-1} b_i \\
\neg q_{m+n} \\
\neg p_{m+n} \\
\neg a_i \lor b_i \lor q_i & i=m,\dots,m+n-1 \\
\neg q_{i+1} \lor q_i & i=m,\dots,m+n-1 \\
a_i \lor q_{i+1} \lor \neg q_i & i=m,\dots,m+n-1 \\
\neg b_i \lor q_{i+1} \lor \neg q_i & i=m,\dots,m+n-1 \\
a_i \lor \neg b_i \lor p_i & i=m,\dots,m+n-1 \\
\neg p_{i+1} \lor p_i & i=m,\dots,m+n-1 \\
\neg a_i \lor p_{i+1} \lor \neg p_i & i=m,\dots,m+n-1 \\
b_i \lor p_{i+1} \lor \neg p_i & i=m,\dots,m+n-1 \\
\bigvee_{i=0}^{m-1} a_i \lor \bigvee_{i=0}^{m-1} b_i \lor d \\
\neg a_i \lor \neg d & i=0,\dots,m-1 \\
\neg b_i \lor \neg d & i=0,\dots,m-1 \\
\neg d \lor q_i & i=0,\dots,m-1 \\
\neg d \lor p_i & i=0,\dots,m-1 \\
d \lor \neg a_i \lor b_i \lor q_i & i=0,\dots,m-1 \\
d \lor \neg q_{i+1} \lor q_i & i=0,\dots,m-1 \\
d \lor a_i \lor q_{i+1} \lor \neg q_i & i=0,\dots,m-1 \\
d \lor \neg b_i \lor q_{i+1} \lor \neg q_i & i=0,\dots,m-1 \\
d \lor a_i \lor \neg b_i \lor p_i & i=0,\dots,m-1 \\
d \lor \neg p_{i+1} \lor p_i & i=0,\dots,m-1 \\
d \lor \neg a_i \lor p_{i+1} \lor \neg p_i & i=0,\dots,m-1 \\
d \lor b_i \lor p_{i+1} \lor \neg p_i & i=0,\dots,m-1 \\
q_0 \lor p_0 \\
\neg q_i \lor \neg p_i \lor q_{i+1} \lor p_{i+1} & i=0,\dots,m+n-1 \\
a_{m+n} \lor b_{m+n} \lor \neg q_i \lor p_i & i=0,\dots,m+n-1 \\ 
\neg a_{m+n} \lor \neg b_{m+n} \lor q_i \lor \neg p_i & i=0,\dots,m+n-1 \\ 
\end{array}
\label{eq:less-than-rule}
\end{equation}
where $d$, $q_i$, and $p_i$ are intermediate Boolean variables.

This rule generates $2m+2n+2$ intermediate Boolean variables and $15m + 11n + 6$ clauses.

\subsection{Reduction Rule for Less-Than-or-Equal-To Expression: $LessEqual(\boldsymbol a, \boldsymbol b)$}
The rule models the expression $\boldsymbol a \le \boldsymbol b$, where $\boldsymbol a$ and $\boldsymbol b$ are signed binary fixed-point numbers. The clauses generated by this rule are shown in Formula~\ref{eq:less-equal-rule}.
\begin{equation}
\begin{array}{cl}
a_{m+n} \lor \neg b_{m+n} \lor \neg a_i & i=0,\dots,m+n-1 \\
a_{m+n} \lor \neg b_{m+n} \lor \neg b_i & i=0,\dots,m+n-1 \\
\neg q_{m+n} \\
\neg p_{m+n} \\
\neg a_i \lor b_i \lor q_i & i=m,\dots,m+n-1 \\
\neg q_{i+1} \lor q_i & i=m,\dots,m+n-1 \\
a_i \lor q_{i+1} \lor \neg q_i & i=m,\dots,m+n-1 \\
\neg b_i \lor q_{i+1} \lor \neg q_i & i=m,\dots,m+n-1 \\
a_i \lor \neg b_i \lor p_i & i=m,\dots,m+n-1 \\
\neg p_{i+1} \lor p_i & i=m,\dots,m+n-1 \\
\neg a_i \lor p_{i+1} \lor \neg p_i & i=m,\dots,m+n-1 \\
b_i \lor p_{i+1} \lor \neg p_i & i=m,\dots,m+n-1 \\
\bigvee_{i=0}^{m-1} a_i \lor \bigvee_{i=0}^{m-1} b_i \lor d \\
\neg a_i \lor \neg d & i=0,\dots,m-1 \\
\neg b_i \lor \neg d & i=0,\dots,m-1 \\
\neg d \lor q_i & i=0,\dots,m-1 \\
\neg d \lor p_i & i=0,\dots,m-1 \\
d \lor \neg a_i \lor b_i \lor q_i & i=0,\dots,m-1 \\
d \lor \neg q_{i+1} \lor q_i & i=0,\dots,m-1 \\
d \lor a_i \lor q_{i+1} \lor \neg q_i & i=0,\dots,m-1 \\
d \lor \neg b_i \lor q_{i+1} \lor \neg q_i & i=0,\dots,m-1 \\
d \lor a_i \lor \neg b_i \lor p_i & i=0,\dots,m-1 \\
d \lor \neg p_{i+1} \lor p_i & i=0,\dots,m-1 \\
d \lor \neg a_i \lor p_{i+1} \lor \neg p_i & i=0,\dots,m-1 \\
d \lor b_i \lor p_{i+1} \lor \neg p_i & i=0,\dots,m-1 \\
a_{m+n} \lor b_{m+n} \lor \neg q_i \lor p_i & i=0,\dots,m+n-1 \\ 
\neg a_{m+n} \lor \neg b_{m+n} \lor q_i \lor \neg p_i & i=0,\dots,m+n-1 \\ 
\end{array}
\label{eq:less-equal-rule}
\end{equation}
where $d$, $q_i$, and $p_i$ are intermediate Boolean variables.

This rule generates $2m+2n+3$ intermediate Boolean variables and $16m+12n+3$ clauses.

\subsection{Reduction Rule for the Multiplier: $Multiplier(\boldsymbol a, \boldsymbol b, \boldsymbol c)$}
The rule models an multiplier that takes signed binary fixed-point numbers $\boldsymbol a$ and $\boldsymbol b$ as inputs, produces the product $\boldsymbol c$. The clauses generated by this rule are shown in Formula~\ref{eq:multiplier-rule}.
\begin{equation}
\begin{array}{cl}
 \neg a_{m+n} \lor b_{m+n} \lor c_{m+n} \\
 a_{m+n} \lor \neg b_{m+n} \lor c_{m+n} \\
 \neg a_{m+n} \lor \neg b_{m+n} \lor \neg c_{m+n} \\
 a_{m+n} \lor b_{m+n} \lor \neg c_{m+n} \\
 \neg a_i \lor \neg b_j \lor T_{i,j} & i=0,\dots,m+n-1, \space j=0,\dots,m+n-1, \space 1 \le i+j \le 2m+n-1 \\
 a_i \lor \neg T_{i,j} & i=0,\dots,m+n-1, \space j=0,\dots,m+n-1, \space 1 \le i+j \le 2m+n-1 \\
 b_j \lor \neg T_{i,j} & i=0,\dots,m+n-1, \space j=0,\dots,m+n-1, \space 1 \le i+j \le 2m+n-1 \\
 \neg d_{i,0} \lor T_{i,0} & i=1,\dots,m+n-1 \\
 d_{i,0} \lor \neg T_{i,0} & i=1,\dots,m+n-1 \\
 \neg d_{m+n,0} \\
 \neg e_{0,i} & i = 1,\dots,m+n-1 \\
 FullAdder(T_{i,j}, d_{i+1,j-1}, e_{i,j}, d_{i,j}, e_{i+1,j}) & i=0,\dots,m+n-1, \space j=1,\dots,m+n-1, \space 1 \le i+j \le 2m+n-1 \\
 \neg d_{m+n,j} \lor e_{m+n,j} & j = 1,\dots,m-1 \\
 d_{m+n,j} \lor \neg e_{m+n,j} & j = 1,\dots,m-1 \\
 \neg c_i \lor d_{i+m-\min\{i+m, m+n-1\}, \min\{i+m, m+n-1\}} & i=0,\dots,m+n-1 \\
 c_i \lor \neg d_{i+m-\min\{i+m, m+n-1\}, \min\{i+m, m+n-1\}} & i=0,\dots,m+n-1 \\
 \neg e_{2m+n-j,j} & j=m,\dots,m+n-1 \\
 \neg a_i \lor \neg b_i & i=0,\dots,m+n-1, \space j=0,\dots,m+n-1, i+j > 2m+n-1 \\
\end{array}
\label{eq:multiplier-rule}
\end{equation}
where $T_{i,j}$, $d_{i,j}$, and $e_{i,j}$ are intermediate Boolean variables.

This rule generates $2m^2+4mn+m+\frac{3}{2}n^2 +\frac{3}{2}n$ intermediate Boolean variables and $13m^2 + 26mn - 4m +7n^2 +2n + 1$ clauses.

\subsection{Reduction Rule for Product Expression: $Product(\{\boldsymbol a_1, \boldsymbol a_2, \cdots, \boldsymbol a_k\}, \boldsymbol c)$}
The rule models the expression $\boldsymbol c = \prod_{i=1}^k \boldsymbol a_i$, where $\boldsymbol a_i$ and $\boldsymbol c$ are signed binary fixed-point numbers. The clauses generated by this rule are shown in Formula~\ref{eq:product-rule}.
\begin{equation}
\begin{array}{cl}
\bigvee_{j=0}^{m+n-1} a_{i,j} \lor d & i=1,\dots,k \\
\bigvee_{j=0}^{m+n-1} a_{i,j} \lor e_i & i = 1,\dots,k \\
\neg a_{i,j} \lor \neg e_i & i=1,\dots,k, \space j=0,\dots,m+n-1 \\
\bigvee_{i=1}^k e_i \lor \neg d \\
\bigvee_{j=0}^{m+n-1} c_j \lor d \\
\neg d \lor \neg c_j & j=0,\dots,m+n-1 \\
d \lor Multiplier(\boldsymbol a_1, \boldsymbol a_2, \boldsymbol b_2) \\
d \lor Multiplier(\boldsymbol b_{i-1}, \boldsymbol a_i, \boldsymbol b_i) & i=3,\dots,k \\
d \lor \neg b_{k,j} \lor c_{j} & j=0,\dots,m+n \\
d \lor b_{k,j} \lor \neg c_{j} & j=0,\dots,m+n \\
\end{array}
\label{eq:product-rule}
\end{equation}
where $d$ and $e_i$ are intermediate Boolean variables.

This rule generates $2(k-1)m^2 + 4(k-1)mn + \frac{3}{2}(k-1)n^2 + 2(k-1)m + \frac{5}{2}(k-1)n + 2k$ intermediate Boolean variables and $13(k-1)m^2 + 26(k-1)mn + 7(k-1)n^2 + (7-3k)m + (3k+1)n +3(k+1)$ clauses.

\subsection{Reduction Rule for Power Expression: $Power(\boldsymbol a, k, \boldsymbol c)$}
The rule models the expression $\boldsymbol c = \boldsymbol a^k$ ($k\in \mathbb{Z}$), where $\boldsymbol a$ and $\boldsymbol c$ are signed binary fixed-point numbers. The clauses generated by this rule are shown in Formula~\ref{eq:power-rule}.
\begin{equation}
\begin{array}{ll}
 \neg(k=0) \lor \neg c_i & i = 0,\dots,m+n, \space i\neq m \\
 \neg(k=0) \lor c_m \\
 \neg(k=1) \lor \neg a_i \lor c_i & i=0,\dots,m+n \\
 \neg(k=1) \lor a_i \lor \neg c_i & i=0,\dots,m+n \\
\neg(k=-1) \lor Product(\{\boldsymbol a, \boldsymbol c\}, \boldsymbol 1) \\
 \neg(|k|>1) \lor Power(\boldsymbol a, \lfloor |k|/2 \rfloor, \boldsymbol d_1) \\
\neg(|k|>1) \lor Product(\{\boldsymbol d_1, \boldsymbol d_1\}, \boldsymbol d_2) \\
 \neg(|k|>1) \lor \neg(|k|\space \text{mod} \space 2 = 1) \lor Product(\{\boldsymbol d_2,\boldsymbol a\}, \boldsymbol d_3) \\
 \neg(k>1) \lor \neg d_{2 + (|k|\space \text{mod} \space 2),i} \lor c_i & i=0,\dots,m+n \\
 \neg(k>1) \lor d_{2 + (|k|\space \text{mod} \space 2),i} \lor \neg c_i & i=0,\dots,m+n \\
 \neg(k<-1) \lor Product(\{\boldsymbol d_{2 + (|k|\space \text{mod} \space 2)}, \boldsymbol c\}, \boldsymbol 1)
\end{array}
\label{eq:power-rule}
\end{equation}
where $\boldsymbol d_i$ is an intermediate variable, $\boldsymbol 1$ is the signed binary fixed-point encoding of number 1.

When $k=0$, this rule generates $0$ intermediate Boolean variables and $m+n+1$ clauses.

When $k=1$, this rule generates $0$ intermediate Boolean variables and $2(m+n)+2$ clauses.

When $k=-1$, this rule generates $2 m^2 +4mn + \frac{3}{2} n^2 + 2m + \frac{5}{2} n +4$ intermediate Boolean variables and $13m^2 + 26mn +m + 7 n^2 + m + 7n + 9$ clauses.

When $k>1$, this rule generates $\lfloor \log_2|k| \rfloor(2m+2n+2)+(\sum_{i=2}^{\lfloor \log_2|k| \rfloor} (|k|\space \text{mod} \space 2^i))(2 m^2 +4mn + \frac{3}{2} n^2 + 2m + \frac{5}{2} n +4)$ intermediate Boolean variables and $\lfloor \log_2|k| \rfloor(3m+3n+1)+(\sum_{i=2}^{\lfloor \log_2|k| \rfloor} (|k|\space \text{mod} \space 2^i))(13m^2 + 26mn +m + 7 n^2 + m + 7n + 9)+(1+ (|k| \space \text{mod} \space 2))(m+n+1)$ clauses.

When $k<-1$, this rule generates $\lfloor \log_2|k| \rfloor(2m+2n+2)+(\sum_{i=2}^{\lfloor \log_2|k| \rfloor} (|k|\space \text{mod} \space 2^i))(2 m^2 +4mn + \frac{3}{2} n^2 + 2m + \frac{5}{2} n +4)$ intermediate Boolean variables and $(\lfloor \log_2|k| \rfloor - 1)(3m+3n+1)+(1+\sum_{i=2}^{\lfloor \log_2|k| \rfloor} (|k|\space \text{mod} \space 2^i))(13m^2 + 26mn +m + 7 n^2 + m + 7n + 9)+(1+ (|k| \space \text{mod} \space 2))(m+n+1) + 2(m+n)$ clauses.

\subsection{Reduction Rule for Absolute Value Expression: $Absolute(\boldsymbol a, \boldsymbol c)$}
The rule models the expression $\boldsymbol c = |\boldsymbol a|$, where $\boldsymbol a$ and $\boldsymbol c$ are signed binary fixed-point numbers. The clauses generated by this rule are shown in Formula~\ref{eq:absolute-rule}.
\begin{equation}
\begin{array}{cl}
\neg c_{m+n} \\
\neg a_i \lor c_i & i = 0,\dots,m+n-1 \\
a_i \lor \neg c_i & i=0,\dots,m+n-1 \\
\end{array}
\label{eq:absolute-rule}
\end{equation}

This rule generates $0$ intermediate Boolean variables and $2m+2n+1$ clauses.

\subsection{Reduction Rule for Floor Expression: $Floor(\boldsymbol a, \boldsymbol c)$}
The rule models the expression $\boldsymbol c = \lfloor \boldsymbol a \rfloor$, where $\boldsymbol a$ and $\boldsymbol c$ are signed binary fixed-point numbers. The clauses generated by this rule are shown in Formula~\ref{eq:floor-rule}.
\begin{equation}
\begin{array}{cl}
\neg c_i & i=0,\dots,m-1 \\
\neg a_{m+n} \lor c_{m+n} \\
a_{m+n} \lor \neg c_{m+n} \\
\bigvee_{i=0}^{m-1} a_i \lor d \\
\neg a_i \lor \neg d & i=0,\dots,m-1 \\
\neg d \lor e \\
a_{m+n} \lor e \\
d \lor \neg a_{m+n} \lor \neg e \\
\neg e \lor \neg a_i \lor c_i & i=m,\dots,m+n-1 \\
\neg e \lor a_i \lor \neg c_i & i=m,\dots,m+n-1 \\
\neg h_m \\
e \lor HalfAdder(a_i, h_i, c_i, h_{i+1}) & i=m,\dots,m+n-1 \\
\end{array}
\label{eq:floor-rule}
\end{equation}

This rule generates $n+3$ intermediate Boolean variables and $2m+8n+7$ clauses.

\subsection{Reduction Rule for Ceil Expression: $Ceil(\boldsymbol a, \boldsymbol c)$}
The rule models the expression $\boldsymbol c = \lceil \boldsymbol a \rceil$, where $\boldsymbol a$ and $\boldsymbol c$ are signed binary fixed-point numbers. The clauses generated by this rule are shown in Formula~\ref{eq:ceil-rule}.
\begin{equation}
\begin{array}{cl}
\neg c_i & i=0,\dots,m-1 \\
\neg a_{m+n} \lor c_{m+n} \\
a_{m+n} \lor \neg c_{m+n} \\
\bigvee_{i=0}^{m-1} a_i \lor d \\
\neg a_i \lor \neg d & i=0,\dots,m-1 \\
\neg d \lor e \\
\neg a_{m+n} \lor e \\
d \lor a_{m+n} \lor \neg e \\
\neg e \lor \neg a_i \lor c_i & i=m,\dots,m+n-1 \\
\neg e \lor a_i \lor \neg c_i & i=m,\dots,m+n-1 \\
\neg h_m \\
e \lor HalfAdder(a_i, h_i, c_i, h_{i+1}) & i=m,\dots,m+n-1 \\
\end{array}
\label{eq:ceil-rule}
\end{equation}

This rule generates $n+3$ intermediate Boolean variables and $2m+8n+7$ clauses.

\subsection{Reduction Rule for Maximum Expression: $Max(\{\boldsymbol a_1, \boldsymbol a_2, \cdots, \boldsymbol a_k\}, \boldsymbol c)$}
The rule models the expression $\boldsymbol c = \max_{i=1}^k \{ \boldsymbol a_i \}$, where $\boldsymbol a_i$ and $\boldsymbol c$ are signed binary fixed-point numbers. The clauses generated by this rule are shown in Formula~\ref{eq:max-rule}.
\begin{equation}
\begin{array}{cl}
LessEqual(\boldsymbol a_i, \boldsymbol c) & i=1,\dots,k \\
EnumerationDomain(\boldsymbol c, \{\boldsymbol a_1, \boldsymbol a_2, \cdots, \boldsymbol a_k\}) \\
\end{array}
\label{eq:max-rule}
\end{equation}

This rule generates $3km+3kn+5k$ intermediate Boolean variables and $(21k+2)m + (17k+2)n + (9k+3)$ clauses.

\subsection{Reduction Rule for Minimum Expression: $Min(\{\boldsymbol a_1, \boldsymbol a_2, \cdots, \boldsymbol a_k\}, \boldsymbol c)$}
The rule models the expression $\boldsymbol c = \min_{i=1}^k \{ \boldsymbol a_i \}$, where $\boldsymbol a_i$ and $\boldsymbol c$ are signed binary fixed-point numbers. The clauses generated by this rule are shown in Formula~\ref{eq:min-rule}.
\begin{equation}
\begin{array}{cl}
LessEqual(\boldsymbol c, \boldsymbol a_i) & i=1,\dots,k \\
EnumerationDomain(\boldsymbol c, \{\boldsymbol a_1, \boldsymbol a_2, \cdots, \boldsymbol a_k\}) \\
\end{array}
\label{eq:min-rule}
\end{equation}

This rule generates $3km+3kn+5k$ intermediate Boolean variables and $(21k+2)m + (17k+2)n + (9k+3)$ clauses.

\subsection{Reduction Rule for Integer Domain: $IntegerDomin(\boldsymbol a, L, R)$}
The rule declares that the domain of $\boldsymbol a$ is defined as $\boldsymbol a \in \mathbb{Z}$ and $L \le \boldsymbol a \le R$, where $\boldsymbol a$ is a signed binary fixed-point number, $L$ and $R$ are constant integers($L<R$). The clauses generated by this rule are shown in Formula~\ref{eq:integer-domain-rule}.
\begin{equation}
\begin{array}{l@{\hspace{-12em}}l}
\neg a_i & i=0,\dots,m-1 \\
\neg (k_1 > 0) \lor \neg a_{m+n} \\
\neg (k_2 < 0) \lor a_{m+n} \\
\neg (L = 0) \lor \neg a_{m+n} \lor \neg a_i & i=m,\dots,m+n-1 \\
\neg (R = 0) \lor a_{m+n} \lor \neg a_i & i=m,\dots,m+n-1 \\
\neg a_i & i=m,\dots,m+n-1, \space 2^{i-m} > \max\{|k_1|,|k_2|\} \\
\neg (L\neq 0 \land |L|< 2^n -1 \land |L-1| \neq 2^{\lceil \log_2 \max\{|L|,|R|\} \rceil }) \lor LessEqual(\boldsymbol L, \boldsymbol a) \\
\neg (R \neq 0 \land |R|< 2^n -1 \land |R +1| \neq 2^{\lceil \log_2 \max\{|L|,|R|\} \rceil } ) \lor LessEqual(\boldsymbol a, \boldsymbol R) \\
\end{array}
\label{eq:integer-domain-rule}
\end{equation}

This rule generates $4m+4n+6$ intermediate Boolean variables and $33m+24n+8+\max\{0,n-1 - \log_2 \lfloor \max\{|L|,|R|\} \rfloor\}$ clauses.

\subsection{Reduction Rule for Real Domain: $RealDomin(\boldsymbol a, L_1, R_1, L_2, R_2)$}
The rule declares that the domain of $\boldsymbol a$ is defined as $\boldsymbol a \in \mathbb{R}$, $L_1 \le \boldsymbol a \le L_2$, and $L_2 < \boldsymbol a < R_2$, where $\boldsymbol a$ is a signed binary fixed-point number, $L_1$, $L_2$, $R_1$ and $R_2$ are constant integers ($\max\{L_1, L_2\} < \max\{R_1,R_2\}$). The clauses generated by this rule are shown in Formula~\ref{eq:real-domain-rule}.
\begin{equation}
\begin{array}{l@{\hspace{-18em}}l}
\neg(L_1 > 0 \lor L_2 \ge 0) \lor \neg a_{m+n} \\
\neg(R_1 < 0 \lor R_2 \le 0) \lor a_{m+n} \\
\neg(L_1 = 0) \lor \neg a_{m+n} \lor \neg a_i & i=0,\dots,m+n-1 \\
\neg(R_1 = 0) \lor a_{m+n} \lor \neg a_i & i=0,\dots,m+n-1 \\
\neg(L_2 = 0 \lor R_2 = 0) \lor \bigvee_{i=0}^{m+n-1} a_i \\
\neg a_i & i=0,\dots,m+n-1, \space 2^{i-m} > \max\{|\max\{L_1,L_2\}|,|\min\{R_1,R_2\}|\} \\
\neg(L_1 \neq 0 \land |L_1|< 2^n) \lor LessEqual(\boldsymbol L_1, \boldsymbol a) \\
\neg(L_2 \neq 0 \land |L_2|< 2^n \land |L_2| \neq 2^{\lceil \log_2 \max\{|\max\{L_1,L_2\}|,|\min\{R_1,R_2\}|\} \rceil}) \lor LessThan(\boldsymbol L_2, \boldsymbol a) \\
\neg(R_1 \neq 0 \land |R_1|< 2^n) \lor LessEqual(\boldsymbol a, \boldsymbol R_1) \\
\neg(R_2 \neq 0 \land |R_2|< 2^n \land |R_2| \neq 2^{\lceil \log_2 \max\{|\max\{L_1,L_2\}|,|\min\{R_1,R_2\}|\} \rceil}) \lor LessThan(\boldsymbol a, \boldsymbol R_2) \\
\end{array}
\label{eq:real-domain-rule}
\end{equation}

This rule generates $62m+46n+20+\max\{0,n-1 - \log_2 \lfloor \max\{|\max\{L_1,L_2\}|,|\min\{R_1,R_2\}|\} \rfloor\}$ clauses and $8m+8n+10$ intermediate Boolean variables.

\subsection{Reduction Rule for Enumeration Domain: $EnumerationDomain(\boldsymbol a, \{\boldsymbol b_1, \boldsymbol b_2, \cdots, \boldsymbol b_k\})$}
The rule declares that the domain of $\boldsymbol a$ is defined as $\boldsymbol a \in \{\boldsymbol b_1, \boldsymbol b_2, \cdots, \boldsymbol b_k\}$, where $\boldsymbol a$ and $\boldsymbol b_i$ are signed binary fixed-point numbers. The clauses generated by this rule are shown in Formula~\ref{eq:enum-domain-rule}.
\begin{equation}
\begin{array}{cl}
\neg a_{j} \lor \neg b_{i,j} \lor d_{i,j} & i=1,\dots,k,\space j=0,\dots,m+n \\
a_j \lor b_{i,j} \lor d_{i,j} & i=1,\dots,k,\space j=0,\dots,m+n \\
a_j \lor \neg b_{i,j} \lor \neg d_{i,j} & i=1,\dots,k,\space j=0,\dots,m+n \\
\neg a_j \lor b_{i,j} \lor \neg d_{i,j} & i=1,\dots,k,\space j=0,\dots,m+n \\
\bigvee_{j=0}^{m+n} \neg d_{i,j} \lor e_i & i=1,\dots,k \\
d_{i,j} \lor \neg e_i & i=1,\dots,k,\space j=0,\dots,m+n \\
\bigvee_{i=1}^k e_i \\
\bigvee_{i=1}^k \neg b_{i,j} \lor a_j & j=0,\dots,m+n \\
\bigvee_{i=1}^k b_{i,j} \lor \neg a_j & j=0,\dots,m+n \\
\end{array}
\label{eq:enum-domain-rule}
\end{equation}
where $e_i$ and $d_{i,j}$ are intermediate Boolean variables.

This rule generates $km+kn+2k$ intermediate Boolean variables and $(5k+2) m+(5k+2)n+6k+3$ clauses.

\subsection{Reduction Rule for the Normalization of Objective Value: $Normalization(\boldsymbol a, \boldsymbol c)$}
The rule models the expression $\boldsymbol c = \boldsymbol a + 2^n$, where $\boldsymbol a$ is a signed binary fixed-point number, and $\boldsymbol b$ is an unsigned binary fixed-point number. The clauses generated by this rule are shown in Formula~\ref{eq:normalization-rule}.
\begin{equation}
\begin{array}{cl}
\begin{matrix}
Complement(\boldsymbol a, (\neg c_{m+n})c_{m+n-1}\cdots c_0) \\
\end{matrix}
\end{array}
\label{eq:normalization-rule}
\end{equation}

This rule generates $m+n+1$ intermediate Boolean variables and $8m+8n+5$ clauses.

\section{Detailed Comparison in Solution Quality}\label{appendix:detailed-results}

{
\setlength{\tabcolsep}{3pt}

\begin{table}[h]\centering\vspace{-1cm}
\caption{Comparison in Solution Quality (Part 1 of 3)}

\begin{tabular}{l c c c c c c c c c c}
\toprule
Optimization Problem & Instance & \multicolumn{8}{c}{The Objective Values of Solutions} \\
\cmidrule(lr){3-10}
& & Optimal Solution & GORED & CPLEX & Gurobi & SCIP & GA & EA & PSO \\
\midrule
Graph Coloring Problem \cite{BeasleyORLibrary1990}   & myciel3 & 4.0 & \textbf{4.0} & \textbf{4.0} & \textbf{4.0} & \textbf{4.0} & \textbf{4.0} & \textbf{4.0} & 7.0 \\
                            & myciel4 & 5.0 & \textbf{5.0} & \textbf{5.0} & \textbf{5.0} & \textbf{5.0} & 11.0 & 6.0 & 15.0 \\
                            & myciel5 & 6.0 & \textbf{6.0} & \textbf{6.0} & \textbf{6.0} & \textbf{6.0} & 32.0 & 23.0 & 32.0 \\
                            & queen5\_5 & 5.0 & \textbf{5.0} & \textbf{5.0} & \textbf{5.0} & \textbf{5.0} & 13.0 & 12.0 & 19.0 \\
                            & queen6\_6 & 7.0 & \textbf{7.0} & \textbf{7.0} & \textbf{7.0} & \textbf{7.0} & 29.0 & 21.0 & 29.0 \\
                            & queen7\_7 & 7.0 & \textbf{7.0} & \textbf{7.0} & \textbf{7.0} & \textbf{7.0} & 44.0 & 35.0 & 44.0 \\
                            & queen8\_8 & 9.0 & \textbf{9.0} & \textbf{9.0} & \textbf{9.0} & \textbf{9.0} & 59.0 & 51.0 & 57.0 \\
                            & anna & 11.0 & \textbf{11.0} & \textbf{11.0} & \textbf{11.0} & \textbf{11.0} & 121.0 & 93.0 & 101.0 \\
                            & david & 11.0 & \textbf{11.0} & \textbf{11.0} & \textbf{11.0} & \textbf{11.0} & 72.0 & 53.0 & 61.0 \\
                            & huck & 11.0 & \textbf{11.0} & \textbf{11.0} & \textbf{11.0} & \textbf{11.0} & 63.0 & 36.0 & 53.0 \\
                            & jean & 10.0 & \textbf{10.0} & \textbf{10.0} & \textbf{10.0} & \textbf{10.0} & 64.0 & 36.0 & 52.0 \\
                            & games120 & 9.0 & \textbf{9.0} & \textbf{9.0} & \textbf{9.0} & \textbf{9.0} & 107.0 & 92.0 & 94.0 \\
                            & miles250 & 8.0 & \textbf{8.0} & \textbf{8.0} & \textbf{8.0} & \textbf{8.0} & 118.0 & 80.0 & 88.0 \\
\midrule
Traveling Salesman Problem \cite{gerhardTSPLIB1991}   & burma14 & 3323.0 & \textbf{3323.0} & \textbf{3323.0} & \textbf{3323.0} & \textbf{3323.0} & 3406.0 & 3406.0 & 4011.0 \\
                                & ulysses16 & 6859.0 & \textbf{6859.0} & \textbf{6859.0} & \textbf{6859.0} & \textbf{6859.0} & 6919.0 & 6983.0 & 7569.0 \\
                                & ulysses22 & 7013.0 & \textbf{7013.0} & \textbf{7013.0} & \textbf{7013.0} & \textbf{7013.0} & 7073.0 & 7808.0 & 8896.0 \\
                                & gr17 & 2085.0 & \textbf{2085.0} & \textbf{2085.0} & \textbf{2085.0} & \textbf{2085.0} & 2158.0 & 2345.0 & 2751.0 \\
                                & gr21 & 2707.0 & \textbf{2707.0} & \textbf{2707.0} & \textbf{2707.0} & \textbf{2707.0} & 2775.0 & 3612.0 & 4382.0 \\
                                & gr24 & 1272.0 & \textbf{1272.0} & \textbf{1272.0} & \textbf{1272.0} & \textbf{1272.0} & 1326.0 & 1721.0 & 1875.0 \\
                                & fri26 & 937.0 & \textbf{937.0} & \textbf{937.0} & \textbf{937.0} & \textbf{937.0} & 1020.0 & 1149.0 & 1244.0 \\
                                & bayg29 & 1610.0 & \textbf{1610.0} & \textbf{1610.0} & \textbf{1610.0} & \textbf{1610.0} & 1644.0 & 2652.0 & 2416.0 \\
                                & bays29 & 2020.0 & \textbf{2020.0} & \textbf{2020.0} & \textbf{2020.0} & \textbf{2020.0} & 2073.0 & 3319.0 & 3244.0 \\
                                & dantzig42 & 699.0 & \textbf{699.0} & \textbf{699.0} & \textbf{699.0} & \textbf{699.0} & 713.0 & 1143.0 & 1011.0 \\
                                & swiss42 & 1273.0 & \textbf{1273.0} & \textbf{1273.0} & \textbf{1273.0} & \textbf{1273.0} & 1288.0 & 2770.0 & 2441.0 \\
\midrule
Capacitated Vehicle Routing Problem \cite{uchoaCVRPLIB2017}   & E-n13-k4 & 247.0 & \textbf{247.0} & \textbf{247.0} & \textbf{247.0} & \textbf{247.0} & 256.0 & \textbf{247.0} & 278.0 \\
                                          & E-n22-k4 & 375.0 & \textbf{375.0} & \textbf{375.0} & \textbf{375.0} & \textbf{375.0} & \textbf{375.0} & 462.0 & 467.0 \\
                                          & E-n23-k3 & 569.0 & \textbf{569.0} & \textbf{569.0} & \textbf{569.0} & \textbf{569.0} & 570.0 & 720.0 & 791.0 \\
                                          & P-n16-k8 & 450.0 & \textbf{450.0} & \textbf{450.0} & \textbf{450.0} & \textbf{450.0} & \textbf{450.0} & 465.0 & 471.0 \\
                                          & P-n19-k2 & 212.0 & \textbf{212.0} & \textbf{212.0} & \textbf{212.0} & \textbf{212.0} & \textbf{212.0} & 236.0 & 266.0 \\
                                          & P-n20-k2 & 216.0 & \textbf{216.0} & \textbf{216.0} & \textbf{216.0} & \textbf{216.0} & \textbf{216.0} & 268.0 & 295.0 \\
                                          & P-n21-k2 & 211.0 & \textbf{211.0} & \textbf{211.0} & \textbf{211.0} & \textbf{211.0} & \textbf{211.0} & 255.0 & 283.0 \\
                                          & P-n22-k2 & 216.0 & \textbf{216.0} & \textbf{216.0} & \textbf{216.0} & \textbf{216.0} & \textbf{216.0} & 259.0 & 276.0 \\
\midrule
Two-Echelon Vehicle Routing Problem \cite{BeasleyORLibrary1990}   & E-n13-k4-1 & 280.0 & \textbf{280.0} & \textbf{280.0} & \textbf{280.0} & \textbf{280.0} & 290.0 & \textbf{280.0} & 306.0 \\
                                          & E-n13-k4-2 & 286.0 & \textbf{286.0} & \textbf{286.0} & \textbf{286.0} & \textbf{286.0} & 296.0 & \textbf{286.0} & \textbf{286.0} \\
                                          & E-n13-k4-3 & 284.0 & \textbf{284.0} & \textbf{284.0} & \textbf{284.0} & \textbf{284.0} & 296.0 & \textbf{284.0} & 290.0 \\
                                          & E-n13-k4-4 & 218.0 & \textbf{218.0} & \textbf{218.0} & \textbf{218.0} & \textbf{218.0} & 246.0 & \textbf{218.0} & \textbf{218.0} \\
                                          & E-n13-k4-5 & 218.0 & \textbf{218.0} & \textbf{218.0} & \textbf{218.0} & \textbf{218.0} & 232.0 & \textbf{218.0} & 232.0 \\
                                          & E-n13-k4-6 & 230.0 & \textbf{230.0} & \textbf{230.0} & \textbf{230.0} & \textbf{230.0} & 238.0 & \textbf{230.0} & 238.0 \\
                                          & E-n13-k4-7 & 224.0 & \textbf{224.0} & \textbf{224.0} & \textbf{224.0} & \textbf{224.0} & 286.0 & \textbf{224.0} & 246.0 \\
                                          & E-n13-k4-8 & 236.0 & \textbf{236.0} & \textbf{236.0} & \textbf{236.0} & \textbf{236.0} & 290.0 & \textbf{236.0} & 252.0 \\
                                          & E-n13-k4-9 & 244.0 & \textbf{244.0} & \textbf{244.0} & \textbf{244.0} & \textbf{244.0} & 252.0 & \textbf{244.0} & 256.0 \\
                                          & E-n13-k4-10 & 268.0 & \textbf{268.0} & \textbf{268.0} & \textbf{268.0} & \textbf{268.0} & 298.0 & \textbf{268.0} & 278.0 \\
                                          & E-n13-k4-11 & 276.0 & \textbf{276.0} & \textbf{276.0} & \textbf{276.0} & \textbf{276.0} & 290.0 & \textbf{276.0} & 290.0 \\
                                          & E-n13-k4-12 & 290.0 & \textbf{290.0} & \textbf{290.0} & \textbf{290.0} & \textbf{290.0} & \textbf{290.0} & \textbf{290.0} & \textbf{290.0} \\
                                          & E-n13-k4-13 & 288.0 & \textbf{288.0} & \textbf{288.0} & \textbf{288.0} & \textbf{288.0} & 292.0 & \textbf{288.0} & 290.0 \\
                                          & E-n13-k4-14 & 228.0 & \textbf{228.0} & \textbf{228.0} & \textbf{228.0} & \textbf{228.0} & 274.0 & \textbf{228.0} & 262.0 \\
                                          & E-n13-k4-15 & 228.0 & \textbf{228.0} & \textbf{228.0} & \textbf{228.0} & \textbf{228.0} & 270.0 & \textbf{228.0} & 232.0 \\
                                          & E-n13-k4-16 & 238.0 & \textbf{238.0} & \textbf{238.0} & \textbf{238.0} & \textbf{238.0} & 274.0 & \textbf{238.0} & \textbf{238.0} \\
                                          & E-n13-k4-17 & 234.0 & \textbf{234.0} & \textbf{234.0} & \textbf{234.0} & \textbf{234.0} & 286.0 & \textbf{234.0} & 246.0 \\
                                          & E-n13-k4-18 & 246.0 & \textbf{246.0} & \textbf{246.0} & \textbf{246.0} & \textbf{246.0} & 290.0 & \textbf{246.0} & 252.0 \\
                                          & E-n13-k4-19 & 254.0 & \textbf{254.0} & \textbf{254.0} & \textbf{254.0} & \textbf{254.0} & 290.0 & \textbf{254.0} & 256.0 \\
                                          & E-n13-k4-20 & 276.0 & \textbf{276.0} & \textbf{276.0} & \textbf{276.0} & \textbf{276.0} & 290.0 & \textbf{276.0} & 280.0 \\

\bottomrule
\end{tabular}
\end{table}
}

\newgeometry{top=2.1cm, bottom=2.1cm, left=1cm, right=1cm}
{
\setlength{\tabcolsep}{3pt}

\begin{table}\centering
\caption{Comparison in Solution Quality (Part 2 of 3)}

\begin{tabular}{l c c c c c c c c c c}
\toprule
Optimization Problem & Instance & \multicolumn{8}{c}{The Objective Values of Solutions} \\
\cmidrule(lr){3-10}
& & Optimal Solution & GORED & CPLEX & Gurobi & SCIP & GA & EA & PSO \\
\midrule
Multidimensional Knapsack Problem \cite{BeasleyORLibrary1990}   & mknap1-1 & 3800.0 & \textbf{3800.0} & \textbf{3800.0} & \textbf{3800.0} & \textbf{3800.0} & \textbf{3800.0} & \textbf{3800.0} & \textbf{3800.0} \\
                                       & mknap1-2 & 8706.1 & \textbf{8706.1} & \textbf{8706.1} & \textbf{8706.1} & \textbf{8706.1} & \textbf{8706.1} & \textbf{8706.1} & \textbf{8706.1} \\
                                       & mknap1-3 & 4015.0 & \textbf{4015.0} & \textbf{4015.0} & \textbf{4015.0} & \textbf{4015.0} & \textbf{4015.0} & \textbf{4015.0} & 3985.0 \\
                                       & mknap1-4 & 6120.0 & \textbf{6120.0} & \textbf{6120.0} & \textbf{6120.0} & \textbf{6120.0} & \textbf{6120.0} & \textbf{6120.0} & 6000.0 \\
                                       & mknap1-5 & 12400.0 & \textbf{12400.0} & \textbf{12400.0} & \textbf{12400.0} & \textbf{12400.0} & \textbf{12400.0} & \textbf{12400.0} & 12360.0 \\
                                       & mknap1-6 & 10618.0 & \textbf{10618.0} & \textbf{10618.0} & \textbf{10618.0} & \textbf{10618.0} & 10604.0 & \textbf{10618.0} & 10531.0 \\
                                       & mknap1-7 & 16537.0 & \textbf{16537.0} & \textbf{16537.0} & \textbf{16537.0} & \textbf{16537.0} & 16452.0 & 16519.0 & 16374.0 \\
                                       & mknap2-11 & 4554.0 & \textbf{4554.0} & \textbf{4554.0} & \textbf{4554.0} & \textbf{4554.0} & \textbf{4554.0} & \textbf{4554.0} & 4527.0 \\
                                       & mknap2-12 & 4536.0 & \textbf{4536.0} & \textbf{4536.0} & \textbf{4536.0} & \textbf{4536.0} & \textbf{4536.0} & \textbf{4536.0} & 4411.0 \\
                                       & mknap2-13 & 4115.0 & \textbf{4115.0} & \textbf{4115.0} & \textbf{4115.0} & \textbf{4115.0} & \textbf{4115.0} & \textbf{4115.0} & 3985.0 \\
                                       & mknap2-14 & 4561.0 & \textbf{4561.0} & \textbf{4561.0} & \textbf{4561.0} & \textbf{4561.0} & \textbf{4561.0} & \textbf{4561.0} & 4505.0 \\
                                       & mknap2-15 & 4514.0 & \textbf{4514.0} & \textbf{4514.0} & \textbf{4514.0} & \textbf{4514.0} & \textbf{4514.0} & \textbf{4514.0} & 3870.0 \\
                                       & mknap2-16 & 5557.0 & \textbf{5557.0} & \textbf{5557.0} & \textbf{5557.0} & \textbf{5557.0} & \textbf{5557.0} & \textbf{5557.0} & 5140.0 \\
                                       & mknap2-17 & 5567.0 & \textbf{5567.0} & \textbf{5567.0} & \textbf{5567.0} & \textbf{5567.0} & \textbf{5567.0} & \textbf{5567.0} & 5331.0 \\
                                       & mknap2-18 & 5605.0 & \textbf{5605.0} & \textbf{5605.0} & \textbf{5605.0} & \textbf{5605.0} & \textbf{5605.0} & \textbf{5605.0} & 5592.0 \\
                                       & mknap2-19 & 5246.0 & \textbf{5246.0} & \textbf{5246.0} & \textbf{5246.0} & \textbf{5246.0} & \textbf{5246.0} & \textbf{5246.0} & 5164.0 \\
                                       & mknap2-20 & 6339.0 & \textbf{6339.0} & \textbf{6339.0} & \textbf{6339.0} & \textbf{6339.0} & \textbf{6339.0} & \textbf{6339.0} & 6005.0 \\
                                       & mknap2-21 & 5643.0 & \textbf{5643.0} & \textbf{5643.0} & \textbf{5643.0} & \textbf{5643.0} & \textbf{5643.0} & \textbf{5643.0} & 4955.0 \\
                                       & mknap2-22 & 6339.0 & \textbf{6339.0} & \textbf{6339.0} & \textbf{6339.0} & \textbf{6339.0} & \textbf{6339.0} & \textbf{6339.0} & 5900.0 \\
                                       & mknap2-23 & 6159.0 & \textbf{6159.0} & \textbf{6159.0} & \textbf{6159.0} & \textbf{6159.0} & \textbf{6159.0} & \textbf{6159.0} & 5862.0 \\

\midrule
Job-Shop Scheduling Problem \cite{TAILLARD1993}   & Ta01 & 1231.0 & \textbf{1231.0} & \textbf{1231.0} & \textbf{1231.0} & \textbf{1231.0} & 1757.0 & 1778.0 & 1549.0 \\
                                 & Ta02 & 1244.0 & \textbf{1244.0} & \textbf{1244.0} & \textbf{1244.0} & \textbf{1244.0} & 1759.0 & 1744.0 & 1688.0 \\
                                 & Ta03 & 1218.0 & \textbf{1218.0} & \textbf{1218.0} & \textbf{1218.0} & \textbf{1218.0} & 1711.0 & 1757.0 & 1554.0 \\
                                 & Ta04 & 1175.0 & \textbf{1175.0} & \textbf{1175.0} & \textbf{1175.0} & \textbf{1175.0} & 1749.0 & 1762.0 & 1590.0 \\
                                 & Ta05 & 1224.0 & \textbf{1224.0} & \textbf{1224.0} & \textbf{1224.0} & \textbf{1224.0} & 1778.0 & 1722.0 & 1663.0 \\
                                 & Ta06 & 1238.0 & \textbf{1238.0} & \textbf{1238.0} & \textbf{1238.0} & \textbf{1238.0} & 1721.0 & 1764.0 & 1611.0 \\
                                 & Ta07 & 1227.0 & \textbf{1227.0} & \textbf{1227.0} & \textbf{1227.0} & \textbf{1227.0} & 1768.0 & 1786.0 & 1629.0 \\
                                 & Ta08 & 1217.0 & \textbf{1217.0} & \textbf{1217.0} & \textbf{1217.0} & \textbf{1217.0} & 1745.0 & 1775.0 & 1593.0 \\
                                 & Ta09 & 1274.0 & \textbf{1274.0} & \textbf{1274.0} & \textbf{1274.0} & \textbf{1274.0} & 1758.0 & 1818.0 & 1705.0 \\
                                 & Ta10 & 1241.0 & \textbf{1241.0} & \textbf{1241.0} & \textbf{1241.0} & \textbf{1241.0} & 1765.0 & 1811.0 & 1533.0 \\
\midrule
Open-Shop Scheduling Problem \cite{TAILLARD1993}   & Ta01 & 193.0 & \textbf{193.0} & \textbf{193.0} & \textbf{193.0} & \textbf{193.0} & \textbf{193.0} & \textbf{193.0} & 196.0 \\
                                  & Ta02 & 236.0 & \textbf{236.0} & \textbf{236.0} & \textbf{236.0} & \textbf{236.0} & \textbf{236.0} & \textbf{236.0} & 245.0 \\
                                  & Ta03 & 271.0 & \textbf{271.0} & \textbf{271.0} & \textbf{271.0} & \textbf{271.0} & \textbf{271.0} & \textbf{271.0} & 272.0 \\
                                  & Ta04 & 250.0 & \textbf{250.0} & \textbf{250.0} & \textbf{250.0} & \textbf{250.0} & \textbf{250.0} & \textbf{250.0} & 255.0 \\
                                  & Ta05 & 295.0 & \textbf{295.0} & \textbf{295.0} & \textbf{295.0} & \textbf{295.0} & \textbf{295.0} & \textbf{295.0} & 301.0 \\
                                  & Ta06 & 189.0 & \textbf{189.0} & \textbf{189.0} & \textbf{189.0} & \textbf{189.0} & \textbf{189.0} & \textbf{189.0} & 209.0 \\
                                  & Ta07 & 201.0 & \textbf{201.0} & \textbf{201.0} & \textbf{201.0} & \textbf{201.0} & \textbf{201.0} & \textbf{201.0} & 204.0 \\
                                  & Ta08 & 217.0 & \textbf{217.0} & \textbf{217.0} & \textbf{217.0} & \textbf{217.0} & \textbf{217.0} & \textbf{217.0} & 229.0 \\
                                  & Ta09 & 261.0 & \textbf{261.0} & \textbf{261.0} & \textbf{261.0} & \textbf{261.0} & \textbf{261.0} & \textbf{261.0} & 272.0 \\
                                  & Ta10 & 217.0 & \textbf{217.0} & \textbf{217.0} & \textbf{217.0} & \textbf{217.0} & \textbf{217.0} & \textbf{217.0} & 221.0 \\
                                  & Ta11 & 300.0 & \textbf{300.0} & \textbf{300.0} & \textbf{300.0} & \textbf{300.0} & 307.0 & 301.0 & 309.0 \\
                                  & Ta12 & 262.0 & \textbf{262.0} & \textbf{262.0} & \textbf{262.0} & \textbf{262.0} & 267.0 & 264.0 & 277.0 \\
                                  & Ta13 & 323.0 & \textbf{323.0} & \textbf{323.0} & \textbf{323.0} & \textbf{323.0} & 345.0 & 337.0 & 347.0 \\
                                  & Ta14 & 310.0 & \textbf{310.0} & \textbf{310.0} & \textbf{310.0} & \textbf{310.0} & 321.0 & 312.0 & 336.0 \\
                                  & Ta15 & 326.0 & \textbf{326.0} & \textbf{326.0} & \textbf{326.0} & \textbf{326.0} & 335.0 & 332.0 & 360.0 \\
                                  & Ta16 & 312.0 & \textbf{312.0} & \textbf{312.0} & \textbf{312.0} & \textbf{312.0} & 323.0 & 321.0 & 361.0 \\
                                  & Ta17 & 303.0 & \textbf{303.0} & \textbf{303.0} & \textbf{303.0} & \textbf{303.0} & 319.0 & 308.0 & 314.0 \\
                                  & Ta18 & 300.0 & \textbf{300.0} & \textbf{300.0} & \textbf{300.0} & \textbf{300.0} & 310.0 & 301.0 & 316.0 \\
                                  & Ta19 & 353.0 & \textbf{353.0} & \textbf{353.0} & \textbf{353.0} & \textbf{353.0} & 363.0 & 366.0 & 389.0 \\
                                  & Ta20 & 326.0 & \textbf{326.0} & \textbf{326.0} & \textbf{326.0} & \textbf{326.0} & 346.0 & 331.0 & 361.0 \\
\midrule
Quadratic Assignment Problem \cite{QAPLIB1991}   & esc16d & 16.0 & \textbf{16.0} & \textbf{16.0} & \textbf{16.0} & \textbf{16.0} & \textbf{16.0} & \textbf{16.0} & \textbf{16.0} \\
                                  & esc16e & 28.0 & \textbf{28.0} & \textbf{28.0} & \textbf{28.0} & \textbf{28.0} & 30.0 & \textbf{28.0} & 32.0 \\
                                  & esc16f & 0.0 & \textbf{0.0} & \textbf{0.0} & \textbf{0.0} & \textbf{0.0} & \textbf{0.0} & \textbf{0.0} & \textbf{0.0} \\
                                  & esc16g & 26.0 & \textbf{26.0} & \textbf{26.0} & \textbf{26.0} & \textbf{26.0} & \textbf{26.0} & \textbf{26.0} & \textbf{26.0} \\
                                  & esc16i & 14.0 & \textbf{14.0} & \textbf{14.0} & \textbf{14.0} & \textbf{14.0} & \textbf{14.0} & \textbf{14.0} & \textbf{14.0} \\
                                  & esc16j & 8.0 & \textbf{8.0} & \textbf{8.0} & \textbf{8.0} & \textbf{8.0} & \textbf{8.0} & \textbf{8.0} & \textbf{8.0} \\
                                  & esc32e & 2.0 & \textbf{2.0} & \textbf{2.0} & \textbf{2.0} & \textbf{2.0} & \textbf{2.0} & \textbf{2.0} & \textbf{2.0} \\
                                  & esc32g & 6.0 & \textbf{6.0} & \textbf{6.0} & \textbf{6.0} & \textbf{6.0} & \textbf{6.0} & \textbf{6.0} & \textbf{6.0} \\

\bottomrule
\end{tabular}
\end{table}

\begin{table}\centering
\caption{Comparison in Solution Quality (Part 3 of 3)}

\begin{tabular}{l c c c c c c c c c c}
\toprule
Optimization Problem & Instance & \multicolumn{8}{c}{The Objective Values of Solutions} \\
\cmidrule(lr){3-10}
& & Optimal Solution & GORED & CPLEX & Gurobi & SCIP & GA & EA & PSO \\
\midrule
Shifted Sphere Function \cite{CEC2005}   & sphere10 & -450.0 & \textbf{-450.0} & \textbf{-450.0} & \textbf{-450.0} & \textbf{-450.0} & \textbf{-450.0} & \textbf{-450.0} & \textbf{-450.0} \\
                                  & sphere20 & -450.0 & \textbf{-450.0} & \textbf{-450.0} & \textbf{-450.0} & \textbf{-450.0} & \textbf{-450.0} & \textbf{-450.0} & \textbf{-450.0} \\
                                  & sphere30 & -450.0 & \textbf{-450.0} & \textbf{-450.0} & \textbf{-450.0} & \textbf{-450.0} & \textbf{-450.0} & \textbf{-450.0} & \textbf{-450.0} \\
                                  & sphere40 & -450.0 & \textbf{-450.0} & \textbf{-450.0} & \textbf{-450.0} & \textbf{-450.0} & \textbf{-450.0} & \textbf{-450.0} & \textbf{-450.0} \\
                                  & sphere50 & -450.0 & \textbf{-450.0} & \textbf{-450.0} & \textbf{-450.0} & \textbf{-450.0} & \textbf{-450.0} & \textbf{-450.0} & \textbf{-450.0} \\
                                  & sphere60 & -450.0 & \textbf{-450.0} & \textbf{-450.0} & \textbf{-450.0} & \textbf{-450.0} & \textbf{-450.0} & \textbf{-450.0} & \textbf{-450.0} \\
                                  & sphere70 & -450.0 & \textbf{-450.0} & \textbf{-450.0} & \textbf{-450.0} & \textbf{-450.0} & \textbf{-450.0} & \textbf{-450.0} & \textbf{-450.0} \\
                                  & sphere80 & -450.0 & \textbf{-450.0} & \textbf{-450.0} & \textbf{-450.0} & \textbf{-450.0} & \textbf{-450.0} & \textbf{-450.0} & \textbf{-450.0} \\
                                  & sphere90 & -450.0 & \textbf{-450.0} & \textbf{-450.0} & \textbf{-450.0} & \textbf{-450.0} & \textbf{-450.0} & \textbf{-450.0} & \textbf{-450.0} \\
                                  & sphere100 & -450.0 & \textbf{-450.0} & \textbf{-450.0} & \textbf{-450.0} & \textbf{-450.0} & \textbf{-450.0} & \textbf{-450.0} & -438.3 \\
\midrule
Shifted Schwefel's Problem \cite{CEC2005}   & schwefel10 & -450.0 & \textbf{-450.0} & \textbf{-450.0} & \textbf{-450.0} & \textbf{-450.0} & \textbf{-450.0} & \textbf{-450.0} & \textbf{-450.0} \\
                                   & schwefel20 & -450.0 & \textbf{-450.0} & \textbf{-450.0} & \textbf{-450.0} & \textbf{-450.0} & -449.1 & \textbf{-450.0} & \textbf{-450.0} \\
                                   & schwefel30 & -450.0 & \textbf{-450.0} & \textbf{-450.0} & \textbf{-450.0} & \textbf{-450.0} & -445.2 & 103.6 & \textbf{-450.0} \\
                                   & schwefel40 & -450.0 & \textbf{-450.0} & \textbf{-450.0} & \textbf{-450.0} & \textbf{-450.0} & -441.4 & 8233.4 & \textbf{-450.0} \\
                                   & schwefel50 & -450.0 & \textbf{-450.0} & \textbf{-450.0} & \textbf{-450.0} & \textbf{-450.0} & -412.6 & 32025.9 & -434.1 \\
                                   & schwefel60 & -450.0 & \textbf{-450.0} & \textbf{-450.0} & \textbf{-450.0} & \textbf{-450.0} & -189.9 & 58776.7 & 954.3 \\
\midrule
Shifted Rosenbrock's Function \cite{CEC2005}   & rosenbrock10 & -390.0 & \textbf{-390.0} & \textbf{-390.0} & \textbf{-390.0} & \textbf{-390.0} & -367.0 & \textbf{-390.0} & -368.3 \\
                                      & rosenbrock20 & -390.0 & \textbf{-390.0} & \textbf{-390.0} & \textbf{-390.0} & \textbf{-390.0} & -318.6 & -380.1 & -364.4 \\
                                      & rosenbrock30 & -390.0 & \textbf{-390.0} & \textbf{-390.0} & \textbf{-390.0} & \textbf{-390.0} & -269.4 & -364.4 & -376.5 \\
                                      & rosenbrock40 & -390.0 & \textbf{-390.0} & \textbf{-390.0} & \textbf{-390.0} & \textbf{-390.0} & -305.7 & -355.8 & -364.9 \\
                                      & rosenbrock50 & -390.0 & \textbf{-390.0} & \textbf{-390.0} & \textbf{-390.0} & \textbf{-390.0} & -178.3 & -344.2 & -304.9 \\
                                      & rosenbrock60 & -390.0 & \textbf{-390.0} & \textbf{-390.0} & \textbf{-390.0} & \textbf{-390.0} & -304.4 & -334.4 & 510.9 \\
                                      & rosenbrock70 & -390.0 & \textbf{-390.0} & \textbf{-390.0} & \textbf{-390.0} & \textbf{-390.0} & -164.1 & -326.8 & -207.2 \\
                                      & rosenbrock80 & -390.0 & \textbf{-390.0} & \textbf{-390.0} & \textbf{-390.0} & \textbf{-390.0} & -105.1 & -315.0 & -245.1 \\
                                      & rosenbrock90 & -390.0 & \textbf{-390.0} & \textbf{-390.0} & \textbf{-390.0} & \textbf{-390.0} & -212.9 & -306.1 & 371.8 \\
                                      & rosenbrock100 & -390.0 & \textbf{-390.0} & \textbf{-390.0} & \textbf{-390.0} & \textbf{-390.0} & -232.3 & -297.7 & -242.6 \\

\bottomrule
\end{tabular}
\end{table}

}

\end{document}